\documentclass[aps,prd,longbibliography,superscriptaddress ,twocolumn,nofootinbib,10pt]{revtex4-2}

\usepackage{graphicx, comment}
\usepackage{amsmath}
\usepackage{amsfonts}
\usepackage{graphicx}
\usepackage{dsfont}
\usepackage{amssymb}
\usepackage{ifthen}
\usepackage{ulem}
\usepackage{color}
\usepackage{float}
\usepackage{physics }
\usepackage{hyperref}
\usepackage{bbold}
\usepackage{lipsum}
\usepackage{nccmath}
\usepackage{xcolor}

\hypersetup{colorlinks=true}

\newcommand{\bkfa}{Ba$_{1-x}$K$_x$Fe$_2$As$_2$ }
\newcommand{\bkfano}{Ba$_{1-x}$K$_x$Fe$_2$As$_2$}
\newcommand{\biJ}{bilinear Josephson }

\newcommand{\btrs}{\text{BTRS}}

\newcommand{\uwv}{\text{wv}}

\begin{document}

\title{
Ultrasound  evidence for multicomponent superconducting order parameter in \bkfa with electron quadrupling phase }

\author{Chris Halcrow}
\email{chalcrow@kth.se}
\affiliation{Department of Physics, KTH Royal Institute of Technology, SE-106 91 Stockholm, Sweden}
\author{Ilya Shipulin}
\affiliation{Institute for Solid State and Materials Physics, Technische Universit\"at Dresden, 01069 Dresden, Germany}
\affiliation{Tsung-Dao Lee Institute \& School of Physics and Astronomy, Shanghai Jiao Tong University, Shanghai 200240, China}
\author{Federico Caglieris}
\affiliation{University of Genoa, Via Dodecaneso 33, 16146 Genoa, Italy}
\affiliation{Consiglio Nazionale delle Ricerche (CNR)-SPIN, Corso Perrone 24, 16152 Genova, Italy}
\affiliation{Leibniz Institute for Solid State and Materials Research, 01069, Dresden, Germany}
\author{Yongwei Li}
\affiliation{Tsung-Dao Lee Institute \& School of Physics and Astronomy, Shanghai Jiao Tong University, Shanghai 200240, China}
\author{Joachim Wosnitza}
\affiliation{Institute for Solid State and Materials Physics, Technische Universit\"at Dresden, 01069 Dresden, Germany}
\affiliation{Dresden High Magnetic Field Laboratory (HLD-EMFL) and Würzburg-Dresden Cluster of Excellence ct.qmat, Helmholtz-Zentrum Dresden-Rossendorf, Dresden, Germany}
\author{Hans-Henning Klauss}
\affiliation{Institute for Solid State and Materials Physics, Technische Universit\"at Dresden, 01069 Dresden, Germany}
\author{Sergei Zherlitsyn}
\affiliation{Dresden High Magnetic Field Laboratory (HLD-EMFL) and Würzburg-Dresden Cluster of Excellence ct.qmat, Helmholtz-Zentrum Dresden-Rossendorf, Dresden, Germany}
\author{Vadim Grinenko}
\email{vadim.grinenko@sjtu.edu.cn}
\affiliation{Tsung-Dao Lee Institute \& School of Physics and Astronomy, Shanghai Jiao Tong University, Shanghai 200240, China}
\author{Egor Babaev}
\email{babaev@kth.se}
\affiliation{Department of Physics, KTH Royal Institute of Technology, SE-106 91 Stockholm, Sweden}
\affiliation{Wallenberg Initiative Materials Science for Sustainability, Department of Physics, KTH Royal Institute of Technology, SE-106 91 Stockholm, Sweden}

\begin{abstract}
Experiments have pointed to the formation of 
the  electron quadrupling condensate in  \bkfa at $x \sim 0.8$.
The state spontaneously breaks time-reversal symmetry and is sandwiched between two critical points, separating it from the broken time-reversal symmetry (BTRS) superconducting state at 
$T_{\rm c}^{U(1)}$ and 
normal-metal state at
$T_{\rm c}^{\rm Z2}$. 
  We report a theory of the acoustic effects spectroscopy 
  of systems with an electron quadrupling phase based on ultrasound-velocity measurements. 
  We show that the experimental results are consistent with BTRS superconductivity at $x \sim 0.8$, fulfilling the necessary condition for the formation of 
  electron quadrupling in \bkfa.
We provide the theoretical basis and the experimental strategy to study the order parameter symmetry of emerging quadrupling condensates in superconductors.  

 \end{abstract}
 
\maketitle

\section{Introduction}

The electron quadrupling condensate is defined as a state whose order parameter 
is composed out of fermionic operators.
In the case of \bkfano, evidence was provided of an order parameter of the type ${\langle c_{\sigma i}c_{\alpha i}c_{\sigma j} ^\dagger c_{\alpha j}^\dagger\rangle}$, where $\alpha,\sigma$ are spin indices and $i,j$ are band, or more generally, indices \cite{Grinenko2021state}.
Such states can appear in systems which, at low temperatures, support a superconducting state that breaks multiple symmetries \cite{Svistunov2015}.
In \bkfa this state was observed at so-called magic doping, where the system is in a paired superconducting state characterised by multiple pairing fields $\langle c_{\sigma i}c_{\alpha i}\rangle$, associated with  multiple gaps,  that break time-reversal symmetry.
The breaking of time-reversal symmetry is expected to be associated with $s+is$ or $s+id$  or similar pairing \cite{Grinenko2017bkfa, Grinenko2020superconductivity} which has a
 doubly degenerate (denoted as $Z_2$) phase difference between the components of the order parameter. Time reversal/complex conjugation takes the system between these two states. At elevated temperatures the long-range order in pairing fields $\langle c_{\sigma i}c_{\alpha i}\rangle$ is lost but the system retains $Z_2$-symmetry-breaking. Such a state is characterized by a four-electron order parameter ${\langle c_{\sigma i}c_{\alpha i}c_{\sigma j} ^\dagger c_{\alpha j}^\dagger\rangle}$. For early theory discussions, see \cite{Bojesen2013time, bojesen2014phase}. Away from ``magic doping", the inter-gap phase locking is singly degenerate and the system is time-reversal symmetric and has only one phase transition.
The evidence for this state at magic doping comes from calorimetric, transport, thermoelectric, and muon-spin rotation probes which all suggest that it exists in a range of temperatures $T_c^{U(1)}<T<T_c^{Z_2}$ \cite{Grinenko2021state,shipulin2023calorimetric}. Below $T_c^{U(1)}$, the system undergoes
another phase transition to a superconducting state, signaling the onset of order at the level of electron pairs ${\langle c_{\sigma i}c_{\alpha i}\rangle}$. 
Recently, Zheng et al. \cite{zheng2024observation} reported the creation of a related bosonic state. 
A mechanism describing the formation of fluctuation-induced composite electronic and bosonic orders has been   studied in \cite{babaev2004superconductor, babaev2004phase, Smiseth2005field, Kuklov2008deconfined, Bojesen2013time, bojesen2014phase, Herland2010phase, Svistunov2015, agterberg2008dislocations, berg2009charge, Radzihovsky2009liquid}. These models also predict vortices carrying a fraction of magnetic flux quantum and such vortices were recently observed in this compound \cite{Iguchi2023}. While the experimental data gathered on \bkfa at $x\approx 0.8$  reveals a set of unprecedented properties, most of the properties of this state remain unexplored.  

One of the powerful methods to detect phase transitions, diagnose new states of matter and get insights into symmetries of the order parameters is ultrasound, which allows one to extract elastic constants of materials \cite{hebel1957nuclear, golding1985observation, muller1986observation, tinkham2004introduction,ghosh2020one, Benhabib2021, Grinenko2021state}. In a conventional one-component superconductor, there is a discontinuous jump in ``compressional" ultrasound modes. This is because compressional strain always couples to the magnitude of the superconducting order parameter squared, $|\psi|^2$. In the other sound modes, there is a continuous change in the response due to higher-order coupling. The acoustic response is usually measured 
in a pulse-echo experiment or 
using resonant ultrasound spectroscopy. 

\begin{figure}
    \centering
    \includegraphics[width=1\linewidth]{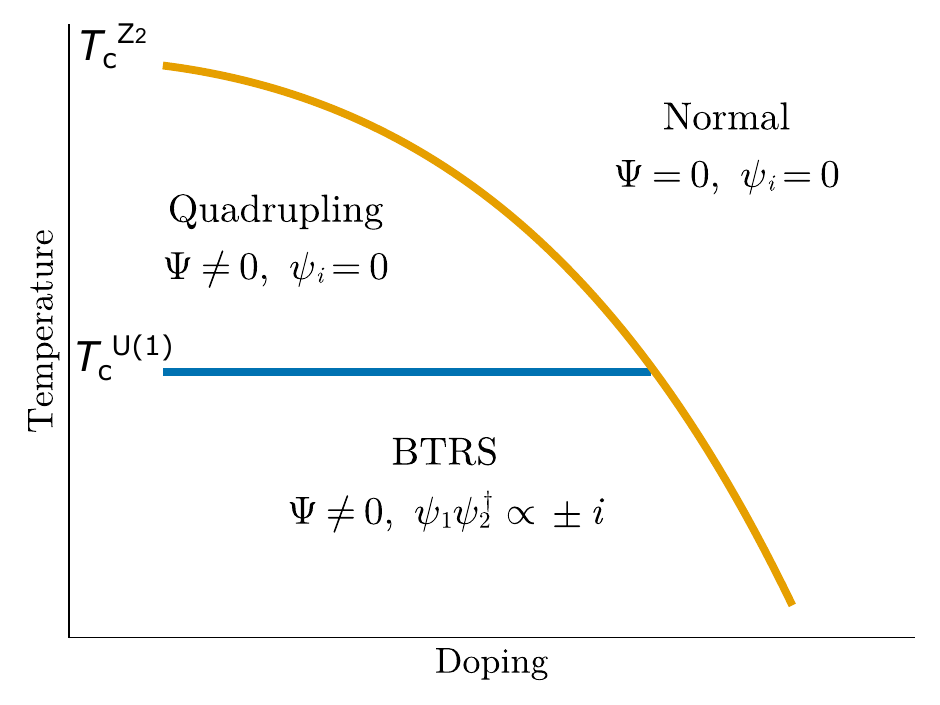}
    \caption{A schematic plot of our phase diagram with BTRS dome and quartic phase constructed according to experimental data~\cite{Grinenko2020superconductivity,Grinenko2021state,shipulin2023calorimetric}}.
    \label{fig:phase_diagram}
\end{figure}

Complex ultrasound responses, such as discontinuous jumps in non-compressional sound modes, are predicted for unconventional superconductors \cite{sigrist2002ehrenfest} and are currently actively searched for in a variety of materials \cite{ghosh2020one, ghosh2021thermodynamic, Benhabib2021}. The responses can be used to get insight into the order-parameter symmetry and structure. In a previous work, we observed an unprecedented type of ultrasound response in \bkfa at doping $x=0.81$ \cite{Grinenko2021state}.
All the other previously mentioned unusual responses occur at a similar doping, $x\approx$ 0.8, and, hence, we will refer to it as the ``magic doping". A schematic phase diagram at magic doping is shown in Fig.~\ref{fig:phase_diagram}. Away from magic doping, the ultrasound response is conventional. However, in the investigated sample with $x=0.81$, there are two very distinct ultrasound singularities occurring
at different temperatures.
First, there is a 
feature in the transverse 
mode in the quadrupling state, $T_c^{U(1)} < T < T_c^{Z_2}$. Second, there are jumps in the ultrasound response in both transverse and longitudinal modes at $T_c^{U(1)}$. These first measurements were performed in zero magnetic field only. To visualize the feature at $T_c^{Z_2}$, the normal-state behavior of the ultrasound velocity was extrapolated using the zero-field data above $T_c^{Z_2}$. This procedure involves a degree of uncertainty in the determination of the continuous features in the temperature dependence of the ultrasound velocity. Therefore, further measurements are required to reconfirm the analyses since, as we will show, these features have strong theoretical implications and deserve careful study and scrutiny.

There is no theory to date to explain the reported ultrasound behavior and deduce whether it is related to the other observed probes \cite{Grinenko2021state,shipulin2023calorimetric}, both near the electron quadrupling
transition at $T_c^{Z_2}$ and the subsequent transition from this state to the superconducting state at $T_c^{U(1)}$. 
In this work, we report both a new set of ultrasound
measurements and a theory of ultrasound probes for the electron quadrupling state.
We show below that the 
ultrasound data of \bkfa with $x \approx 0.75-0.8$ supports the existence of the multicomponent superconducting state. However, to understand the observed jump in the transverse mode $c_{66}$ mode, one is required to assume an $s$+i$d_{\rm xy}$ order parameter or an $s$+i$s$ one with extra factors such as symmetry-breaking strain fields. 
The latter response can originate, e.g., from the enhanced nematic susceptibility along the [110] direction observed recently at low temperatures in the vicinity of magic doping~\cite{hong2022elastoresistivity}. However, the anomaly at $T_c^{Z_2}$ was not resolved within the error bars of the measurements, but the data indicate a possible anomaly at $T_c^{Z_2}$ in the longitudinal mode, favoring an $s$+i$s$ order parameter at this doping level. In a more general context, the presented theoretical work contributes towards deciphering aspects of the momentum-space symmetries of pairing and quadrupling symmetries.

\section{Results}

\subsection{Experimental results}

\begin{figure*}[]
    \centering
    \includegraphics[width=1.0\linewidth]{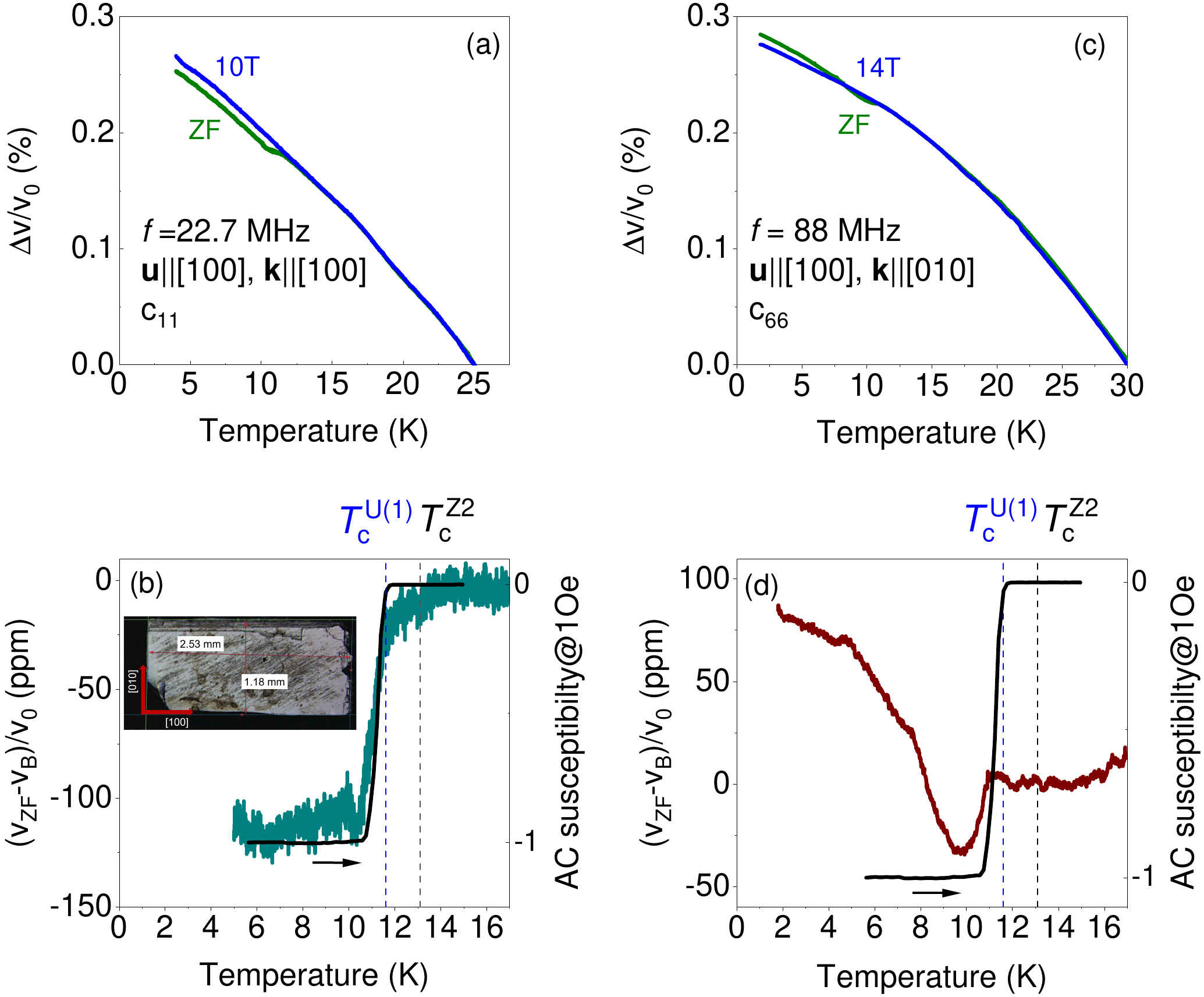}
    \caption{
    Experimental results demonstrating the change of relative sound velocity as a function of temperature for \bkfa  with doping $x\approx0.78$, for a
     longitudinal $c_{11}$ and (c) transverse $c_{66}$ acoustic modes. The measurements were done using a transit acoustic signal (zero echo) at zero field (ZF) and filed applied along the $c$-axis. Temperature dependence of the relative change of the sound velocity (b) for the longitudinal $c_{11}$ and (d) transverse $c_{66}$ acoustic modes (left) with subtracted in-field data and AC magnetic susceptibility (right) measured at $B$ = 1 Oe with $f$ = 417 Hz applied along the $c$-axis. 
    }
    \label{fig:data1}
\end{figure*}

\begin{figure*}[]
    \centering
    \includegraphics[width=1\linewidth]{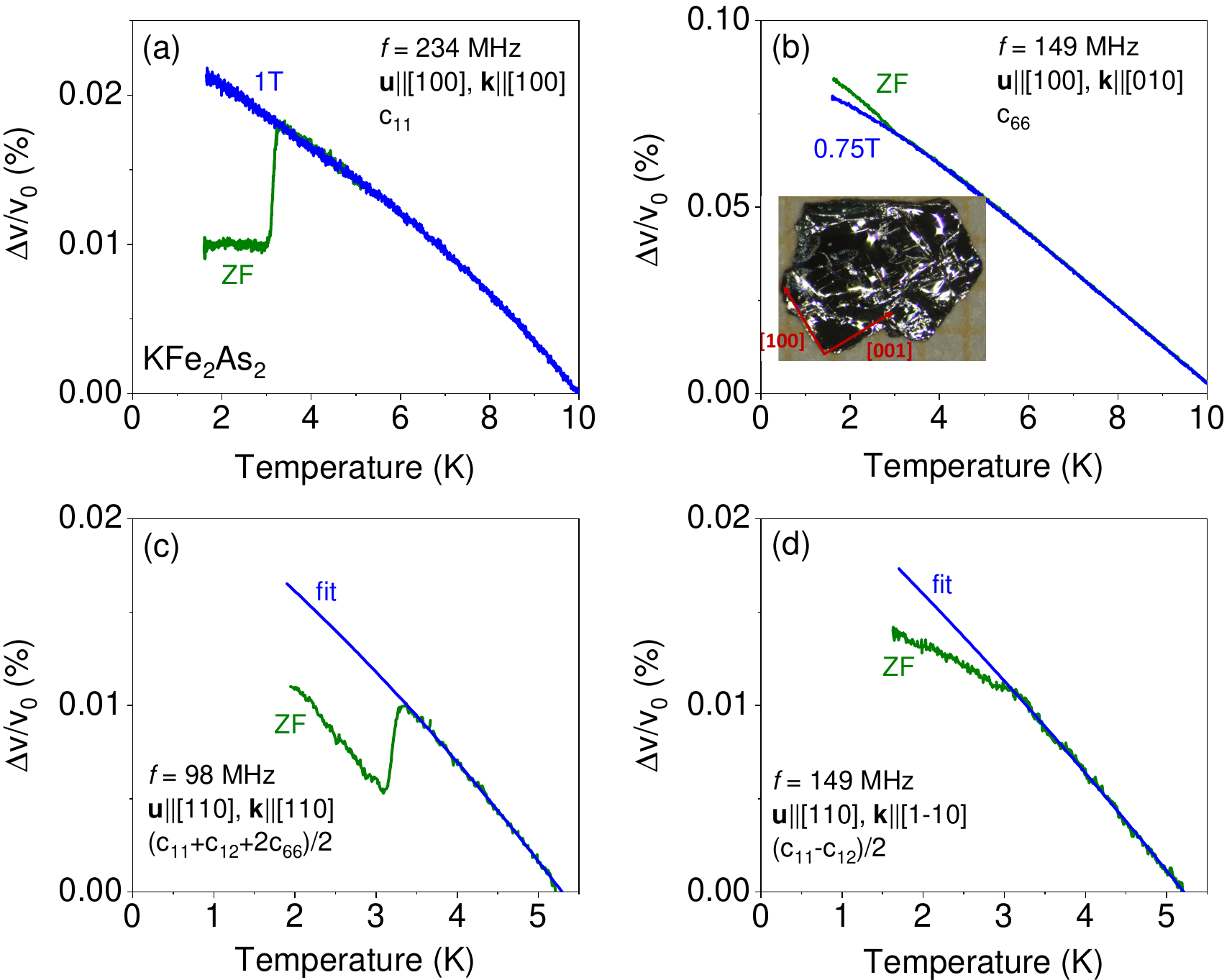}
    \caption{Temperature dependence of the relative change of the sound velocity of KFe$_2$As$_2$ for (a) the longitudinal $c_{11}$, (b) transverse $c_{66}$, (c) longitudinal $(c_{11}+c_{12}+2c_{66})/2$, and (d) transverse $(c_{11}-c_{12})/2$ acoustic modes. The measurements were done using a transit acoustic signal (zero echo) at zero field (ZF) and filed applied along the $c$-axis for the sound propagation along the [100] direction. Only zero-field measurements were performed for the sound propagation along the [110] direction.}
    \label{fig:data2}
\end{figure*}

In this work, we performed ultrasound measurements using two new single crystals, with $x\approx 0.78$ and 1. The first one is with optimal doping for the quartic state with $T_{\rm c}^{\rm U(1)} = 11.6$ K (Fig.~\ref{fig:data1}b), which corresponds to the doping level $x \approx0.78$ according to the previously established phase diagram~\cite{Grinenko2020superconductivity}. Previous systematic $\mu$SR, spontaneous Nernst effect and specific heat studies indicate that the samples with this doping level have $T_{\rm c}^{Z_2} > T_{\rm c}^{\rm U(1)}$ with $T_{\rm c}^{Z_2}  \sim 13$ K 
\cite{Grinenko2021state,shipulin2023calorimetric, Baertl2025}. The second is a stochiometric KFe$_2$As$_2$ sample far away from the broken time-reversal symmetry (BTRS) dome. The measurements were performed using a pulse-echo method. The experimental procedure is described in the Methods section. The photographs of the samples are shown in Figs.~\ref{fig:data1}b and ~\ref{fig:data2}b. 
The direction of the sound wave propagation was along the longest sample side. 

The sample thickness of the single crystal containing the quadrupling state 
was about 50 $\mu$m. This is thicker than previous measurements \cite{Grinenko2021state} but still thin compared to what is typically used for ultrasound experiments on single crystals. This choice is dictated by technical challenges in obtaining thicker homogeneous samples with magic doping. This sample thickness limits the prospect of obtaining optimal ultrasound signals due to possible interference effects. To minimize such interference, all measurements were performed using short-duration zero echoes (accepting only the first-coming signal). This procedure significantly minimizes possible effects of interference. In this study, we restrict ourselves only to qualitative discussions of the character of anomalies and
 do not perform any quantitative analyses of jump heights. 
 
The reference KFe$_2$As$_2$ sample single crystal is 200 $\mu$m thick. This allowed us to obtain much better data quality for KFe$_2$As$_2$. We also took special care in orienting the ultrasound propagation direction with respect to crystallographic directions. The KFe$_2$As$_2$ sample was oriented using the Laue method. However, this method gave inconclusive results for the small crystal with $x\approx 0.78$. Therefore, the orientation was verified using polarized Raman spectroscopy, where the assignment is straightforward (see Methods section).

\begin{figure}[]
    \centering
    \includegraphics[width=1\linewidth]{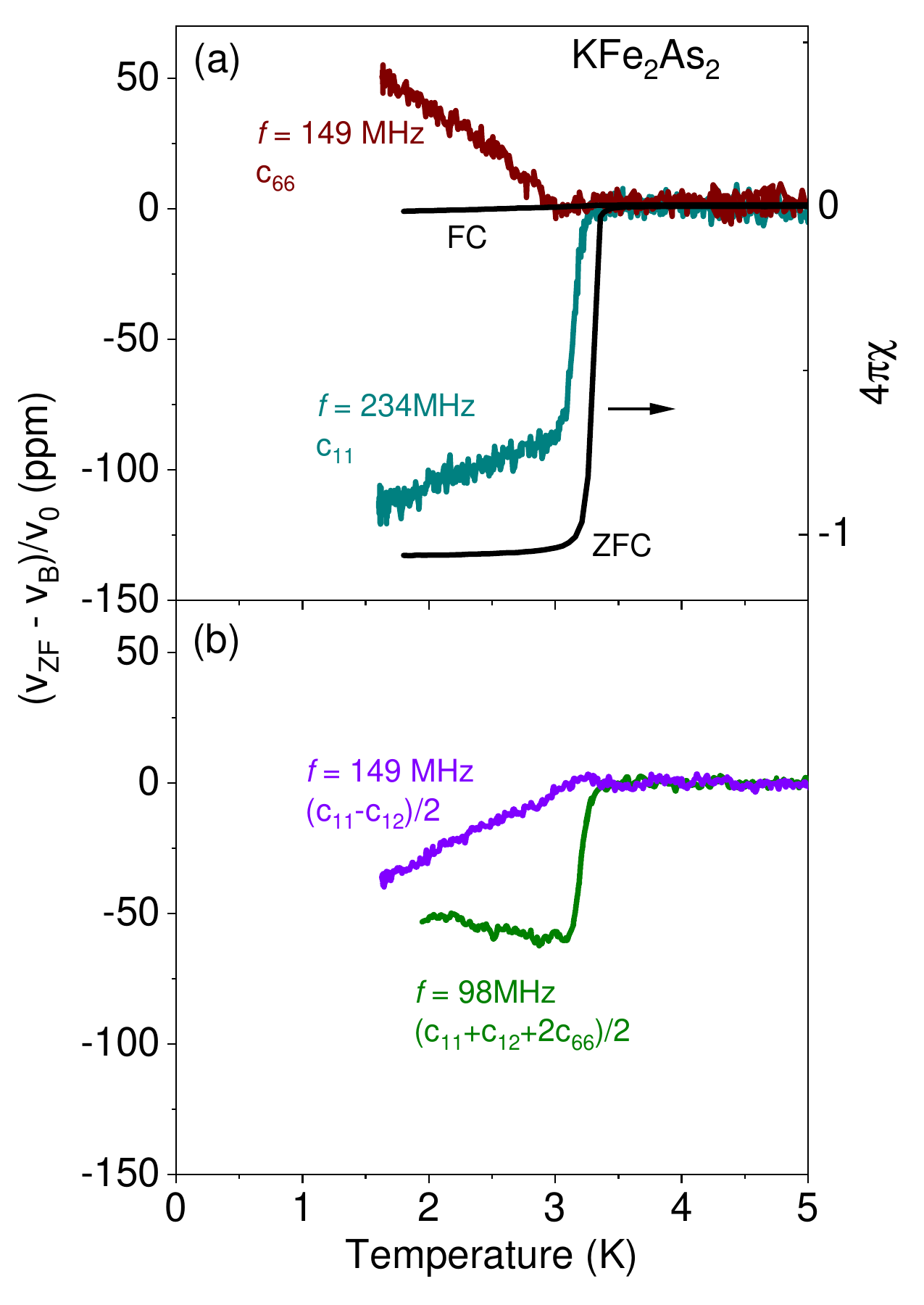}
    \caption{Temperature dependence of the relative change of the sound velocity of KFe$_2$As$_2$ with subtracted normal state contribution for the data shown in Fig.~\ref{fig:data2} for (a) the longitudinal $c_{11}$, and transverse $c_{66}$ acoustic modes (left) and DC magnetic susceptibility (right) measured at $B$ = 5 Oe applied along the $c$-axis and (b) for the longitudinal $(c_{11}+c_{12}+2c_{66})/2$ and transverse $(c_{11}-c_{12})/2$ acoustic modes.
    }
    \label{fig:data3}
\end{figure}

For the sample with quadrupling phase, due to the small sample size, we performed only measurements of the longitudinal compression 
($c_{\rm 11}$) and the transverse ``$B_{2g}$" ($c_{\rm 66}$) shear mode, using as-grown crystal edges. For the reference sample, with a larger sample size, we could measure the longitudinal ``$A_{1g}$" [$(c_{11} + c_{22} + 2c_{66})/2$] and transverse ``$B_{1g}$" [$(c_{\rm 11} - c_{\rm 12})/2$] modes in addition to the $c_{\rm 11}$ and $c_{\rm 66}$ modes. 
In this study, we performed ultrasound measurements in zero and high magnetic fields strong enough to suppress $T_c^{U(1)}$ and $T_c^{Z_2}$ well below the zero-field velues. The results in zero and high magnetic field applied along the $c$-axis are shown in Fig.~\ref{fig:data1} for $x \approx 0.78$, and in Figs.~\ref{fig:data2} and~\ref{fig:data3} for the KFe$_2$As$_2$ sample. For both samples, we observed jumps at $T_c^{U(1)}$ in the sound velocity of the longitudinal acoustic modes. For the sample with $x \approx 0.78$, there is a possible kink close to $T_c^{Z_2} \sim 13$ K 
[Fig.~\ref{fig:data1}b].  However, we did not observe any resolvable features above the superconducting transition in the $c_{\rm 66}$ mode [Fig.~\ref{fig:data1}d.] At $T_c^{U(1)}$, the behaviour of the $c_{66}$ mode is very different for the two samples. The sample with the quadrupling phase shows a jump-like feature close to $T_c^{U(1)}$ [Fig.~\ref{fig:data1}d]. The size of the anomaly is rather large, about 100 times larger than, for instance, the one measured for Sr$_2$RuO$_4$~\cite{Benhabib2021}. In contrast, the reference sample shows a kink at $T_c^{U(1)}$ for this mode [Fig.~\ref{fig:data3}a]. We also did not observe any feature above $T_c^{U(1)}$ in the longitudinal $(c_{11} + c_{22} + 2c_{66})/2$ and transverse $(c_{\rm 11} - c_{\rm 12})/2$ modes and there is no resolvable jump in $(c_{\rm 11} - c_{\rm 12})/2$ at $T_c^{U(1)}$ in the reference sample [Figs.~\ref{fig:data2}c,~\ref{fig:data2}d, and~\ref{fig:data3}a] in agreement with previous measurements \cite{Grinenko2021split}.

\subsection{Theoretical formalism} \label{sec:GLmodel}

The question we address here is: Does the electron quadrupling condensation show itself in the form of singularities in the ultrasound responses?
In the quadrupling phase, there is no long-range ordering bilinear in electronic fields  (i.e., no order in the superconducting gap/order parameter fields).
Furthermore, the mechanism for the formation of the quadrupling state requires fluctuations and is beyond the BCS mean-field approximation \cite{babaev2004superconductor,Bojesen2013time,bojesen2014phase,maccari2022effects,Grinenko2021state,shipulin2023calorimetric}.
Nonetheless, as discussed in models with related kinds of orders \cite{Kuklov2006decoinfined,Svistunov2015}, the resulting phase diagrams with electron quadrupling can, a posteriori, be approximately described by using the ``second" mean-field approximation. This just means a more general approximant involving (non-independent) order parameters of both superconducting and quadrupling order that phenomenologically describe observed broken symmetries.
Following this approach, we will introduce a ``quadrupling" order parameter $\Psi$, alongside the two-component order parameters for the superconducting state ($\psi_1, \psi_2$). In the case of \bkfa, which breaks time-reversal symmetry, $\Psi$ should share the same symmetry as $\psi_1 \psi_2^\dagger$, or, in terms of fermionic creation and annihilation operators, $\Psi \propto$ $\langle c_1c_1c_2^\dagger c_2^\dagger\rangle$. In particular, we require that $\Psi$ is gauge invariant and $\Psi + \Psi^\dagger$ is time-reversal symmetric.

We will now develop a minimal model, which can reproduce the experimental phase diagram. In order of decreasing temperature, the phase diagram consists of normal, electron quadrupling, and BTRS superconducting phases. The model is constructed from a two-component superconducting order parameter $(\psi_1, \psi_2)$ and the quadrupling order parameter $\Psi$. The three phases can be described by the field values of the order parameters (OPs) in them: normal ($\psi_i=\Psi=0$), quadrupling ($\psi_i = 0, \Psi \neq 0$) and BTRS superconducting phase ($\psi_1, \psi_2\ne 0, \text{Im }\psi_1\psi_2^\dagger \neq 0$). A schematic plot of our phase diagram is 
shown in Fig.~\ref{fig:phase_diagram}.

The free energy must be real and time-reversal-symmetric. A Ginzburg-Landau (GL) model which satisfies all these requirements is
\begin{align} \label{eq:free_energy}
&\mathcal{F}_V = -\frac{a(T)}{2}\left( |\psi_1|^2 + |\psi_2|^2 \right) + \frac{b}{4}\left( |\psi_1|^4 + |\psi_2|^4 \right) \nonumber \\
&- A_i(T) \Psi_i^2 + A_r \Psi_r^2 + \frac{B_1}{2} \left(\Psi_r^4 + \Psi_i^4\right) + B_2\Psi_r^2\Psi_i^2 \nonumber \\
&+ c(\psi_1\psi_2^\dagger +\psi_1^\dagger\psi_2)^2 + \frac{\gamma}{4} \left( \Psi \psi_1 \psi_2^\dagger +\Psi^\dagger \psi_1^\dagger \psi_2  \right) \, ,
\end{align}
where $\Psi = \Psi_r + i \Psi_i$. The coefficients of each term are parameters of the model. We use small Latin letters for terms involving just the superconducting OPs, capital Latin letters for terms involving just the quadrupling OP and $\gamma$ for the mixed term. In the quadrupling phase, the ground state of $\Psi$ will be two-fold degenerate with $\Psi = \pm i|\Psi_0|$.
The superconducting and BTRS phase transitions are controlled by the coefficients $a(T)$ and $A_i(T)$, respectively. For simplicity, we consider the following temperature dependence of the coefficients:
\begin{align}
a(T) &= \alpha_{SC}(T
^{\rm U(1)} - T), \\ 
A_i(T) &= \alpha_{\btrs}(T
^{\rm Z2} - T) \, .
\end{align}
where $T
^{\rm U(1)}$ and $T
^{\rm Z2}$  are characteristic constants that coincide with critical temperatures $T_{\rm c}
^{\rm U(1)}$ and $T_{\rm c}
^{\rm Z2}$ in the simplest models.
In these approximations, we aim to reproduce the morphology of the phase diagram of Ba$_{1-x}$K$_x$Fe$_2$As$_2$, which is sufficient for our goal of describing the ultrasound response qualitatively  {Note that \bkfa has more than two bands, and more general models with a higher number of fields are also considered \cite{Grinenko2021state}, some comparative discussion between two- and three-component models can be found in \cite{Garaud2022effective}}.

To calculate the ultrasound response we need to couple the order parameters to the strain of the crystal lattice. We do so following the works on superonductors \cite{ sigrist2002ehrenfest, luthi2007physical}.
The strain energy is written in terms of the strain tensor $u_{i,j} = 1/2(\partial u_i/ \partial x_j + \partial u_j /\partial x_i)$, where $u_i$ is the displacement vector of the underlying crystal lattice. The strain can be labelled by the irreducible representations of the $D_{4h}$ lattice symmetry group of Ba$_{1-x}$K$_x$Fe$_2$As$_2$. The combinations $u_{x,x}+u_{y,y}$ and $u_{z,z}$ transform as $A_{1g}$, $u_{x,x} - u_{y,y}$ transforms as $B_{1g}$, $u_{x,y}$ transforms as $B_{2g}$ and the pair $(u_{x,z},u_{y,z})$ transform as $E_g$. The six independent terms in the elastic energy are given by the six products of these strains that transform as $A_{1g}$. The elastic constants are usually written in Voigt notation with two indices. Using this notation, the strain energy is given by
\begin{align} \label{eq:strain}
\mathcal{F}_S = &\frac{c_{11}+c_{12}}{2}\left( u_{x,x} + u_{y,y}\right)^2 + c_{13}(u_{x,x}+u_{y,y})u_{z,z} \nonumber \\
& + c_{33}u_{z,z}^2 +\frac{c_{11}-c_{12}}{2}\left( u_{x,x} - u_{y,y}\right)^2  \nonumber \\
&+ c_{44}\left( u_{x,z}^2 + u_{y,z}^2 \right) + c_{66}u_{x,y}^2 \, .
\end{align}
This is sometimes written in full tensor notation as
\begin{equation} \label{eq:tensornotation}
    \mathcal{F}_S = \tfrac{1}{2}c_{ijkl}u_{i,j}u_{k,l} \, .
\end{equation}
The experimental data is obtained for sound modes which are ``in plane". Hence, from now on, we only consider strains in the $x$-$y$ plane and neglect any strains involving the $z$-coordinate.

The OPs couple to strain, which ultimately leads to the ultrasound response. The coupling depends on the symmetry of the order parameter. Since we consider a mechanism for which the quadrupling OP $\Psi$ has the same symmetry as $\psi_1\psi_2^\dagger$, the symmetry of $\psi_1$ and $\psi_2$ uniquely specifies the symmetry of all OPs. 

We have two goals: First, to determine how the quadrupling order parameter couples to ultrasound, and second, how this probe can be used to determine the OP symmetries. The leading candidates for the superconducting
OP symmetry of
\bkfa at magic doping are
$s+is$ and $s+id$ states. The analysis of the polarization of the spontaneous magnetic fields detected in $\mu$SR experiments \cite{Grinenko2020superconductivity} favors the interpretation in terms of the $s+is$ states.
However, there is currently not enough certainty about the microscopic details to establish a precise model for spontaneous magnetic fields. They are sensitive to details \cite{benfenati2020magnetic}, including the nature of the magnetic-field-induced disorder and domain-wall structure. 

We are further guided by the fact that there is a non-zero response in the $c_{66}$, or $B_{2g}$, ultrasound mode at $T_C^{U(1)}$. In a simple GL model, this can only be non-zero if some combination of the superconducting order parameters transforms like $B_{2g}$. We list how the different possible symmetries combine in Table \ref{tab:producttable}. Only three combinations contain a copy of $B_{2g}$: $A_{1g}\otimes B_{2g}, A_{2g}\otimes B_{1g}$ and $E_{g}\otimes E_{g}$. In common terminology, these order parameters are $s+d$, $f+d$, and a vector $d$ wave, respectively. In our framework, the first two models are indistinguishable, and so we will focus on the $s+d$ model.

Overall, we consider below three different OP symmetries: $(s,s)$, $(s,d_{xy})$, and vector $(d_{xz},d_{yz})$. These are representative of the case for which the two superconducting OPs transform as $(A_{1g}, A_{1g}), (A_{1g}, B_{2g})$ and $E_g$. Note that the ultrasound response is similar for nodal and nodeless $s$-wave models.

The coupling terms that enter the free energy, $\mathcal{F}_C$, differ for the different OP symmetries. We consider all terms which are second-order in the OP (counting $\Psi$ as quadratic). Then there are four terms that couple to strain. They are:
\begin{equation} \label{eq:quadratics}
|\psi_1|^2 + |\psi_2|^2, \; |\psi_1|^2 - |\psi_2|^2, \; \psi_1\psi^\dagger_2 + \psi^\dagger_1 \psi_2\, ,\; \Psi + \Psi^\dagger .
\end{equation}
We will also include coupling to the higher-order term $|\Psi|^2$.
These terms couple to different strains depending on the OP symmetry. We will denote the coupling coefficients between OPs and strain as $\delta_i$. 

$(s,s)$ OP symmetry: all terms couple to the $A_{1g}$ strain. The free-energy term, which couples strain and the OPs is given by
\begin{align} \label{eq:sscouple}
&\mathcal{F}_C^{s,s} = \big[\delta_1(|\psi_1|^2 + |\psi_2|^2) + \delta_2( |\psi_1|^2 - |\psi_2|^2)+\\
&\delta_3(\psi_1\psi^\dagger_2 +  \psi^\dagger_1 \psi_2) +\tfrac{\delta_4}{2} (\Psi + \Psi^\dagger) + \delta_5 |\Psi|^2\big](u_{x,x}+u_{y,y}) \, . \nonumber
\end{align}

$(s,d_{xy}) $ OP symmetry: the mixed bilinears transform as $B_{2g}$. Hence, they couple to $u_{x,y}$, giving the coupling free energy
\begin{align} \label{eq:sdcouple}
&\mathcal{F}_C^{s,d} = \big[\delta_1(|\psi_1|^2 + |\psi_2|^2) + \delta_2( |\psi_1|^2 - |\psi_2|^2) \nonumber \\ &+ \delta_5 |\Psi|^2\big] (u_{x,x}+u_{y,y}) \nonumber \\
&+ \big[\delta_3(\psi_1\psi^\dagger_2 +  \psi^\dagger_1 \psi_2) +\tfrac{\delta_4}{2} (\Psi + \Psi^\dagger)\big]u_{x,y}
\end{align}

$(d_{xz},d_{yz})$ OP symmetry: the simplest vector OP that transforms like the $E_g$ irrep couples to strain as follows
\begin{align} \label{eq:ddcouple}
\mathcal{F}_C^{d,d} = &[\delta_1(|\psi_1|^2 + |\psi_2|^2) + \delta_5 |\Psi|^2](u_{x,x} + u_{y,y}) \nonumber \\
&+ \delta_2( |\psi_1|^2 - |\psi_2|^2) (u_{x,x}-u_{y,y}) \nonumber \\
&+ \big[\delta_3(\psi_1\psi^\dagger_2 +  \psi^\dagger_1 \psi_2) +\tfrac{\delta_4}{2} (\Psi + \Psi^\dagger)\big]u_{x,y} \, .
\end{align}

In all three cases, the strain-OP coupling free energy can be written as
\begin{equation} \label{eq:Gammadef}
    \mathcal{F}_C = \Gamma_{ij}(\psi,\Psi)u_{i,j} \, ,
\end{equation}
where  $\Gamma_{ij}$ is a matrix of functions depending on the OPs.

We have found the free energy for our theory, including strain coupling. We now develop a rather general theory of the ultrasound response for a class of theories, including ours. We consider a model with order parameters $\Pi_a$, symmetric strain tensor $u_{i,j}$, and linear strain coupling. The total free energy can be written as
\begin{equation}
    \mathcal{F} = V(\Pi) + \tfrac{1}{2}c_{ijkl}u_{i,j}u_{k,l} + \Gamma_{ij}(\Pi)u_{i,j} \, ,
\end{equation}
where $V$ is the free energy of just the OP. For the models considered in this paper: $V=\mathcal{F}_V$ from  \eqref{eq:free_energy}, $c_{ijkl}$ is defined by \eqref{eq:strain} and \eqref{eq:tensornotation}, and $\Gamma$ is defined by \eqref{eq:Gammadef} (which depends on the symmetry of the OPs). The free energy has the solution $(\Pi^0, u^0)$, which satisfies the static equations of motion
\begin{align} \label{eq:EoM}
    \frac{\partial}{\partial \Pi_a}\left( V - \tfrac{1}{2}\Gamma_{ij}c_{ijkl}^{-1}\Gamma_{kl} \right) \bigg\rvert_{\Pi=\Pi^0} = 0, \\
    u^0_{i,j} = c^{-1}_{ijkl}\Gamma_{kl}(\Pi^0). 
\end{align}
Naively, the tensor $c_{ijkl}$ does not have a unique inverse. However, since it is symmetric in $i \leftrightarrow j$ and $k \leftrightarrow l$, it does have a unique inverse with this same symmetry.

We are interested in perturbations around the ground state solution $\Pi^0, u^0_{i,j}$. We denote these as $\Pi_a = \Pi^0_a + \eta_a$ and $u = u^0 + u^\uwv$. One must be careful here, and quotient out gauge transformations. We can do this by choosing the perturbations to be gauge invariant $\eta$. We will not be explicit here, as the details depend on whether the system is in the superconducting or quartic phase. The free energy for the perturbations is
\begin{align}
    \mathcal{F}_2 = \tfrac{1}{2}  \left(V_{ab} + \Gamma_{ij,ab} \partial_j u^0_i \right)\eta_a \eta_b \nonumber \\ + \tfrac{1}{2} c_{ijkl}u^\uwv_{i,j}u^\uwv_{k,l} +  \eta_a  \Gamma_{ij,a}  u_{i,j}^\uwv \, ,
\end{align}
where $f,a = \frac{\partial f}{\partial \Pi_a }\rvert_{\Pi^0}$ for any function $f$. The equations of motion for the perturbations are  
\begin{align}
    \tau_0 \frac{\partial \eta_a}{\partial t} + (V_{ab} + \Gamma_{ij,ab}\partial_j u_i^0)\eta_b + \Gamma_{ij,a}u_{i,j}^\uwv = 0\\
    \rho \ddot{u}^{\text{wv}}_i - \partial_j\left( c_{ijkl}u_{k,l}^\uwv + \eta_a \Gamma_{ij,a} \right) = 0 \, ,
\end{align}
where $\tau_0$ is a phenomenological constant \cite{sigrist2002ehrenfest,luthi2007physical} that controls relaxation time. The equations have solutions
\begin{equation} \label{eq:wavesol}
    \eta_a = A_a e^{ik_i x_i - i \omega t}, \quad u_i^{\text{wv}} = U_{i} e^{ik_i x_i - i \omega t} \, .
\end{equation}
where $\boldsymbol{U}$ and $\boldsymbol{k}$ are the amplitude and wavevector of the ultrasound wave. The ansatz \eqref{eq:wavesol} gives the dispersion relation
\begin{equation} \label{eq:nonldisp}
    \rho \omega^2 U_{i} - c_{ijkl}k_jk_l U_{k} + \Gamma_{ij,a}\tilde{V}^{-1}_{ab} \Gamma_{kl,b}k_j k_l U_{k} = 0 \, ,
\end{equation}
where
\begin{equation}
    \tilde{V}_{ab} = \left(V_{ab} + \Gamma_{ij,ab}\partial_j u_i^0 - i \omega \tau_0 \delta_{ab}\right) \, .
\end{equation}

Different sound modes then correspond to different choices of $\boldsymbol{k}$ and $\boldsymbol{U}$ in the dispersion relation. We are particularly interested in three modes, the longitudinal, transverse ($B_{1g}$), and the $B_{2g}$ mode. The longitudal wave corresponds to the choice $\boldsymbol{U} = (1,0,0)$ and $\boldsymbol{k} = (k_{11},0,0)$, the transverse wave to $\boldsymbol{U} = (1,1,0)/\sqrt{2}$ and $\boldsymbol{k} = (k_T,-k_T,0)/\sqrt{2}$, and the $B_{2g}$ mode to $\boldsymbol{U} = (1,0,0)$ and $\boldsymbol{k} = (0,k_{66},0)/\sqrt{2}$. Substituting these into the dispersion relation \eqref{eq:nonldisp}, we can find $k_{11/T/66}(\omega)$ and then the sound velocity is given by
\begin{align}
 v_{11/T/66} = \frac{\omega}{
 \text{Re }k_{11/T/66}(\omega) } \, .
\end{align}

For analytic results, we take the large $c$, small $\tau_0$ limit. In this limit the rescaled and renormalized change in the sound velocity, relative to the normal-state sound velocity $v^0$, is
\begin{equation} \label{eq:analyticexpression}
    \Delta\tilde{v} = c^0 \frac{v - v^0}{v^0} = -\frac{1}{2} \Gamma_{ij,a}\tilde{V}^{-1}_{ab} \Gamma_{kl,b}U_{i}\hat{k}_j U_{k}\hat{k}_l \, ,
\end{equation}
where $c^0$ is the probed elastic coefficient for each mode. For the longitudinal mode, $c^0_{11} = c_{11}$, for the transverse mode $c^0_T = (c_{11} - c_{12})/2$, and for the $B_{2g}$ mode $c^0_{66} = c_{66}$. The normal sound velocity is closely related, given by $v^0_{11/T/66} = \sqrt{ \rho / c^0_{11/T/66}}$.

\subsection{Theoretical results}

To begin, we investigate an analytically tractable toy model without coupling between superconducting and quadrupling order parameters. The advantage of this model is that we can investigate it analytically and
inspect the roles played by some of the terms.  
Here, the free energy 
is given by \eqref{eq:free_energy} with $\gamma=0$ and 
\begin{align}
a(T) &= \alpha_{SC}(T_{\rm c}^{\rm U(1)} - T) \\, 
A_i(T) &= \alpha_{\btrs}(T_{\rm c}^{\rm Z2} - T) \, ,
\end{align}
and $T_{\rm c}^{\rm U(1)} < T_{\rm c}^{\rm Z2}$. We consider the case for which there is no \biJ term. 

The quadrupling phase occurs when $A_i > 0$ but $a<0$. The superconducting order parameters are zero in this phase and, if $A_r>0$, the only non-zero order parameter is $\Psi_i$, equal to
\begin{equation}
    \Psi_i^2 = \frac{A_i}{B_1} \, .
\end{equation}
The ultrasound response in the quadrupling phase for the $(s,s)$  model is
\begin{align}
\Delta \tilde{v}_{11} = \delta_4^2 D_4 -\frac{\delta_5^2}{2B_1} \, , \quad \Delta \tilde{v}_T =\Delta \tilde{v}_{66} = 0,
\end{align}
where we define
\begin{equation}
D_4(T) = \frac{A_i(T) B_2 }{4 A_i(T) A_r B_2+4 A_r^2 B_1} \, .
\end{equation}
The only non-zero response is in the $c_{11}$ mode. In contrast, the $(s,d_{xy})$ model in the quadrupling phase gives the response
\begin{align}
\Delta \tilde{v}_{11} = -\frac{\delta_5^2}{2B_1}, \quad \Delta \tilde{v}_T =0, \quad \Delta \tilde{v}_{66} = \delta_4^2 D_4\, .
\end{align}
The non-zero response in the transverse mode is linear in $T$ for small $A_i$ (equivalently, near the transition), as seen by a Taylor expansion:
\begin{equation}
D_4 \sim  -\frac{\alpha_{\btrs} B_2  (T-T_{\btrs})}{4 A_r^2 B_1} \, .
\end{equation}
Hence, there is a linear slope in the ultrasound response. Finally the $(d_{xz}, d_{yz})$ OP, in the quadrupling phase, has the response
\begin{align}
\Delta \tilde{v}_{11} = -\frac{\delta_5^2}{2B_1}, \quad \Delta \tilde{v}_{T} = 0, \quad  \Delta \tilde{v}_{66} =\delta_4^2 D_4 \, .
\end{align}
Overall, a non-zero response in the transverse mode is present for the $(s, d_{xy})$  and $(d_{xz}, d_{yz})$ models.

At the superconducting transition, the superconducting OPs become nonzero. We assume that $c>0$, so that the superconducting order parameters have broken time-reversal symmetry. Since they are not coupled to the quadrupling phase, we can find the analytic expression for the solution:
\begin{equation}
    \psi_1^2 = -\psi_2^2 = \frac{a}{b-2c} \, .
\end{equation}
Note that in the superconducting phase of the decoupled model there are four degenerate ground states, meaning that the symmetry is broken to a group with an extra $Z_2$ symmetry: $U(1)\times Z_2 \times Z_2$. This deficiency of the toy model will be fixed when we include a non-zero coupling $\gamma$.

Having found the ground-state solutions, we substitute them into equation \eqref{eq:analyticexpression} to find the ultrasound response, which depends on the chosen OP symmetry. In the three cases we consider, the ultrasound responses in the superconducting phase are
\begin{align*}
(s,s): \,\, &\Delta \tilde{v}_{11} = \delta_4^2D_4 - \frac{2 \delta_1^2}{b-2 c}-\frac{2 \delta_2^2}{b+2 c}-\frac{\delta_3^2}{8 c} -\frac{\delta_5^2}{2B_1},\\
&\Delta \tilde{v}_T = \Delta \tilde{v}_{66} = 0 . \\
(s,d_{xy}): \,\, &\Delta \tilde{v}_{11} = -\frac{2 \delta_1^2}{b-2 c}-\frac{2 \delta_2^2}{b+2 c} -\frac{\delta_5^2}{2B_1},\\
&\Delta v_T = 0, \\
&\Delta \tilde{v}_{66} = \delta_4^2D_4-\frac{\delta_3^2}{8c}  .\\
(d_{xz},d_{yz}):\,\, &\Delta \tilde{v}_{11} = -\frac{2 \delta_1^2}{b-2 c} -\frac{\delta_5^2}{2B_1},\\
&\Delta \tilde{v}_T = -\frac{2 \delta_2^2}{b+2 c}, \\
&\Delta \tilde{v}_{66} = \delta_4^2 D_1  - \frac{\delta_3^2}{8c}.
\end{align*}
These results are nontrivial. The most important fact is that there are jumps in the transverse sound mode, $c_{66}$, in both the $(s,d)$ and $(d,d)$ models but not in the $(s,s)$ model. Hence, the jump at the superconducting transition in the transverse mode, which is clearly seen in the experimental data in fig.~\ref{fig:data1}, cannot be described by this toy $(s,s)$ model.

So far, we have modeled the quadrupling order parameter as a complex field, which has two fluctuating modes. Instead, one could model it as an ``imaginary" order parameter, which only has one fluctuating mode and transforms like the imaginary part of $\Psi$. If this was the case, the model is similar and can be described by the free energy Eq.~\eqref{eq:free_energy}, but with any term proportional to $\Psi_r$ equal to zero, 
so that $A_r = B_2=0$. This affects the ultrasound response, especially in the quadrupling phase, where $D_4$ now vanishes. In this case, there would be no observable response in the transverse modes in the quadrupling phase.

We now present results for a more realistic model, which includes coupling between $\Psi$ and $\psi$. That is, with $\gamma\ne 0$. The free energy for the order parameters is given in equation \eqref{eq:free_energy} with
\begin{align}
(b, A_r, c, B_1, B_2,\gamma) = (1, 0.4, 0.2, 1, -0.1,-0.2) \, .
\end{align}
Due to the nonzero $\gamma$ term, there are no explicit formulae for the order parameters in each phase, though we can find them as a series expansion in $\gamma$. The results to first order are
\begin{align}
    \text{Quadrupling: } &\Psi^2 = -A_i/B_1, \psi = 0 \\ 
    \text{SC: } &\Psi = i\left(\sqrt{A_i/B_1} + \gamma \frac{a}{8 A_i(b - 2c)}\right), \nonumber \\
    &\psi_1\psi_2^\dagger = i\left( \frac{a}{b-2c} -\gamma\frac{\sqrt{A_i}}{2\sqrt{B_1}(b - 2c)} \right) \, . \nonumber
\end{align}
The coupling between strain and the OPs depends on the OP symmetry and is given by Eqs.~\eqref{eq:sscouple}-\eqref{eq:ddcouple} with $\delta_i = 1, i=1-4$ and $\delta_5 = 0.7$ (since $\delta_5$ is the coefficient of a higher-order term) and the phase transition temperatures are controlled by
\begin{align}
    A_i = T
    ^{\rm Z2} - T = 2 - T, \\ 
    a = T
    ^{\rm U(1)} - T = 1 - T \, .
\end{align}

\begin{figure*}
    \centering
    \includegraphics[width=1\linewidth]{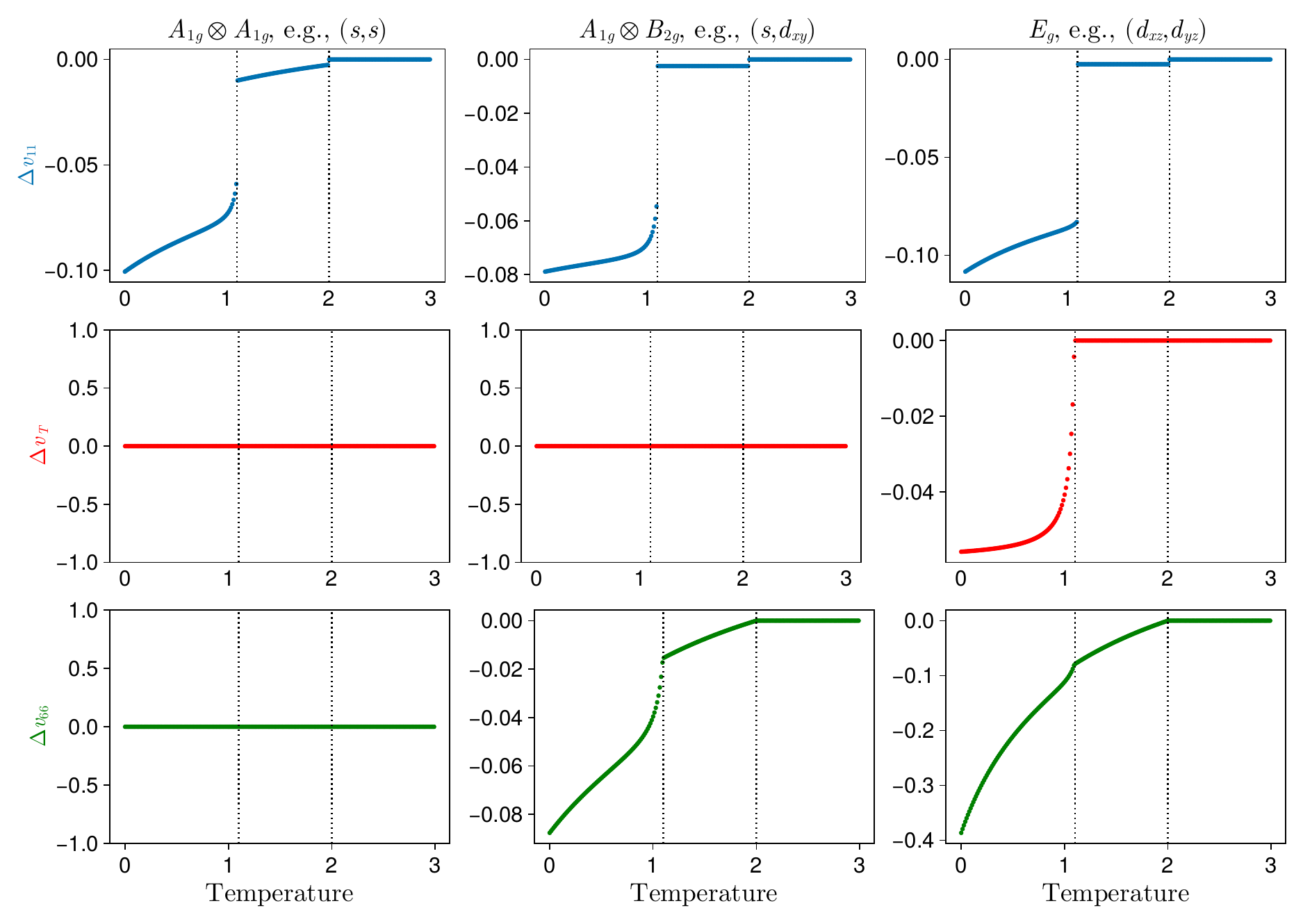}
    \caption{The ultrasound response in the longitudinal, transverse $B_{1g}$ [ $(c_{11}-c_{12})/2$], and transverse $B_{2g}$  $c_{66}$ mode for the three OP cases considered, with the free energy defined in \eqref{eq:free_energy}. The $s, d_{xy}$ model reproduces: a linear change with temperature in the transverse mode below the quadrupling transition $T=T_c^{Z_2}\approx 2$, a minor response in the longitudinal mode at the quadrupling transition, and jumps in both modes at the superconducting transition $T=T_c^{U(1)}\approx 1$. }
    \label{fig:US_response_fav}
\end{figure*}

The results for the $(s,s)$, $(s,d_{xy})$, and $(d_{xz},d_{yz})$ models are shown in Fig.~\ref{fig:US_response_fav}. The coupling between superconducting and quadrupling OPs smooth out the ultrasound responses but still produces anomalies of the scale seen in the experimental data. The $(s,s)$ model (first column in Fig.~\ref{fig:US_response_fav}) has a linear response in the quadrupling state and a jump at the superconducting transition for the longitudinal mode. There is no response for the transverse $(c_{11}-c_{12})/2$ and $c_{66}$ mode. The $(s,d)$ model has a linear response in the $c_{66}$ mode and a small signal in the longitudinal $c_{11}$ mode {Note that this implies there is also a small signal in the  $(c_{11}+c_{12}+2c_{66})/2$ mode. This mode was measured in a previous experiment \cite{Grinenko2021state}, which shows no strong signal.} in the quadrupling phase. At the superconducting transition, the (s,d) model gives a jump in both modes. Again, there is no response in $(c_{11} - c_{12})/2$. The $(d,d)$ model has a jump in the longitudinal mode at the quadrupling phase transition and a jump in both transverse modes at the superconducting transition.

The experimental data suggest that there is a possible weak signal in the longitudinal mode ($c_{11}$), no resolvable response in $c_{66}$ in the quadrupling phase, and jumps in $c_{11}$ and $c_{66}$ at the superconducting transition. Taking these data alone would be consistent with the ($s$,$d_{\rm xy}$) model if the linear response below $T_{\rm c}^{\rm Z2}$ is too weak to be resolved in the present set of data. Another possibility is that the ($s$,$s$) model is coupled to symmetry-breaking strain along the [110] direction discussed below. This strain can be a consequence of the diverging nematic susceptibility at low temperatures reported recently \cite{hong2022elastoresistivity}. In this case, any defect or even ultrasound waves propagating along the [110] direction could induce strong enough strain to produce the observed jump in $c_{66}$ at $T_{\rm c}^{\rm U(1)}$ [Fig.~\ref{fig:data1}(d)]. This scenario would also be consistent with the missing kink at $T_{\rm c}^{\rm Z2}$ in $c_{66}$.

\begin{figure}[h!]
    \centering
 \includegraphics[width=1\linewidth]{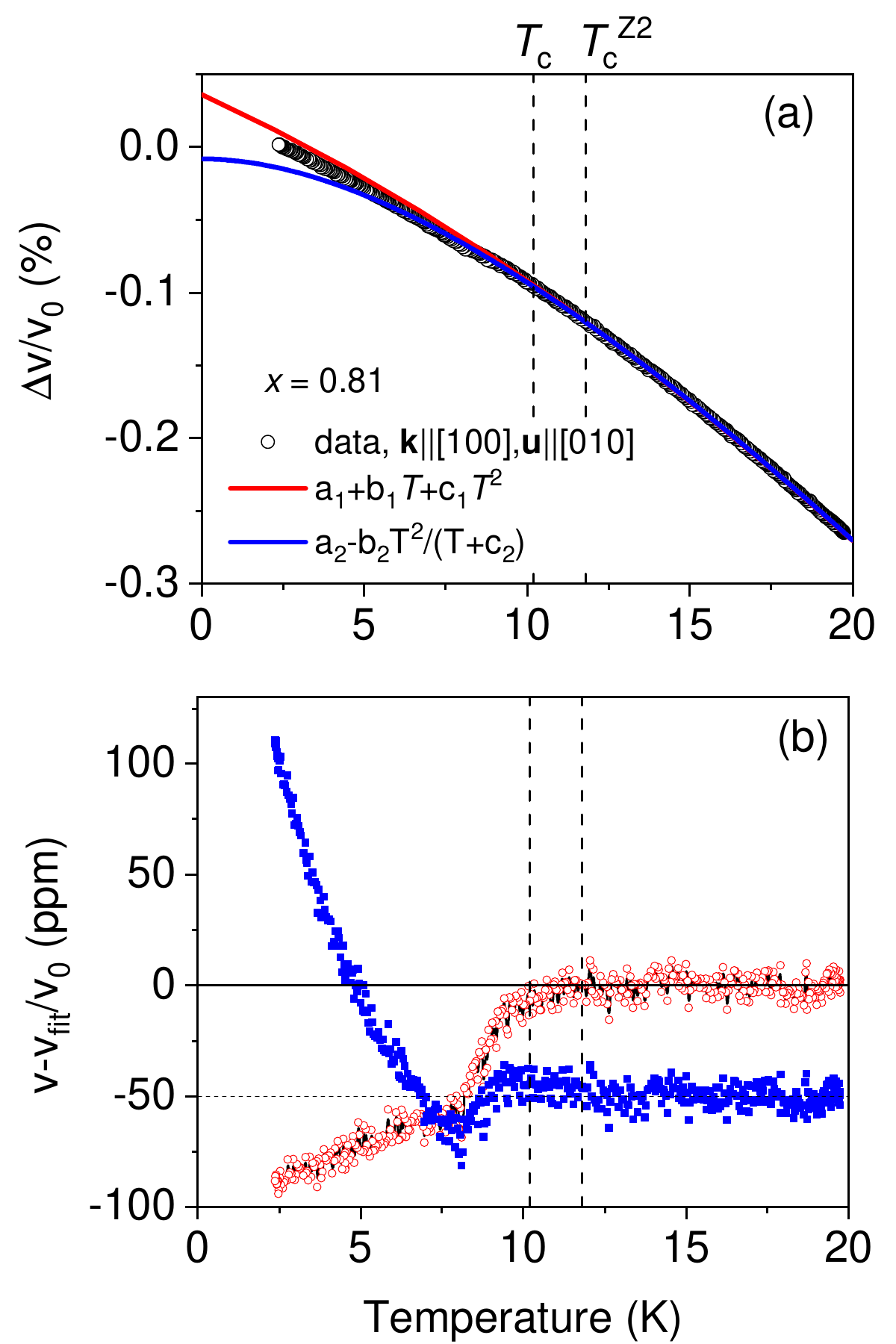}
   \caption{A demonstration that the modeling of the normal-state background field can cause an apparent kink in the ultrasound data.  (a) Temperature dependence of the relative change of the sound velocity for the transverse $(c_{\rm 11} - c_{\rm 12})/2$ acoustic modes for the \bkfa sample with $x = 0.81$ from Ref.~\cite{Grinenko2021state}. The measurements were done using a transit acoustic signal (zero echo) at zero field (ZF). (b) Temperature dependence of the relative change of the sound velocity with the subtracted background using the fitting curves shown in panel (a). The anomaly at $T_{\rm c}^{\rm Z2}$ is sensitive to the fitting procedure.}
    \label{fig:NPfit}
\end{figure}

Previously, we reported a kink-like feature at $T_{\rm c}^{\rm Z2}$ and jump at $T_{\rm c}$ in another transversal mode, namely $(c_{\rm 11} - c_{\rm 12})/2$, for a sample with a quartic state with the slightly different doping level $x = 0.81$~\cite{Grinenko2021state}. We note that these measurements were performed in zero field only and, therefore, the normal-state background was approximated using the temperature dependence above $T_{\rm c}$. This procedure is not well defined, and the kink at $T_{\rm c}^{\rm Z2}$ may be eliminated by adjusting the fit parameters as demonstrated in Fig.~\ref{fig:NPfit}. On the other hand, the jump at $T_{\rm c}$ is more prominent and qualitatively independent of the fitting procedure above $T_{\rm c}$. However, in the previous study, a possible mixing of the [110] and [100] directions cannot be excluded since it was not verified with a reliable method such as Raman spectroscopy. Given these caveats, the previous ultrasound data cannot be reliably used to constrain potential theoretical models. Therefore, further studies are necessary to elucidate the behavior of the $(c_{\rm 11} - c_{\rm 12})/2$ and $(c_{11} + c_{22} + 2c_{66})/2$ modes at magic doping. 

The vector OP $(d_{xz}, d_{yz})$ produces jumps at the superconducting and quadrupling transitions but has no ultrasound response in the transverse mode $(c_{11}-c_{12})/2$ in the quadrupling phase. As we will see near the end of Section \ref{sec:swave}, higher-order terms can produce a weak signal in the quadrupling phase. Hence the experimental data can also, in principle, be described using an OP with this vector symmetry.

\subsection{Analysis of $s$-wave models}
\label{sec:swave}

Momentum-space symmetry of the order parameters in \bkfa remains a subject of discussion.
Initially, several experiments were interpreted in favor of a $d$-wave order parameter in KFe$_2$As$_2$ ($x$ = 1) including thermal conductivity and specific heat \cite{reid2012universal,abdel2013evidence}. However, as mentioned above, the $\mu$SR data favors the scenario that the order parameter is $s$ wave \cite{Grinenko2020superconductivity} at doping $x\approx 0.8$. Recent ARPES data at $x$ = 1 is 
also consistent with an $s_{\pm}$ order parameter \cite{wu2022nodal}. Near optimal doping ($x\approx 0.4$), ARPES \cite{richard2009angle, Cai2021} 
and thermal-conductivity \cite{luo2009quasiparticle} data suggest that the order parameter is $s$-wave and isotropic. Below optimal doping, the gap becomes anisotropic and develops extrema \cite{reid2016doping}, though it is still  typically thought to be $s$ wave. However, our considerations suggest that there is no ultrasound response in the transverse sound modes for the simplest $s+is$  order parameter.
This is inconsistent with the experimental data indicating a linear response in the quadrupling phase, then a jump at $T_{\rm c}^{\rm U(1)}$. Hence, we now explore possible modifications to the $s$ wave theory, which might produce the desired non-trivial response. In this subsection, we always assume that the superconducting OP transforms as $(s,s)$ and the quadrupling OP is $s$ wave.

First let us consider the possibility of nematicity, so that the lattice symmetry changes in the quadrupling state.
The original lattice has $D_{4h}$ symmetry. 
Experimental data from \cite{hong2022elastoresistivity} suggest there is some enhanced nematic susceptibility consistent with the proximity of a ``[110]" nematic critical point close to $x = 0.8$. This stretches the square-like original lattice into a diamond shape, breaking the $D_{4h}$ symmetry to $C_{2h}$. Hence, $90^\circ$ rotations (which form a $C_4$ subgroup) are broken to $180^\circ$ rotations.

\begin{figure}
    \centering
    \includegraphics[width=\linewidth]{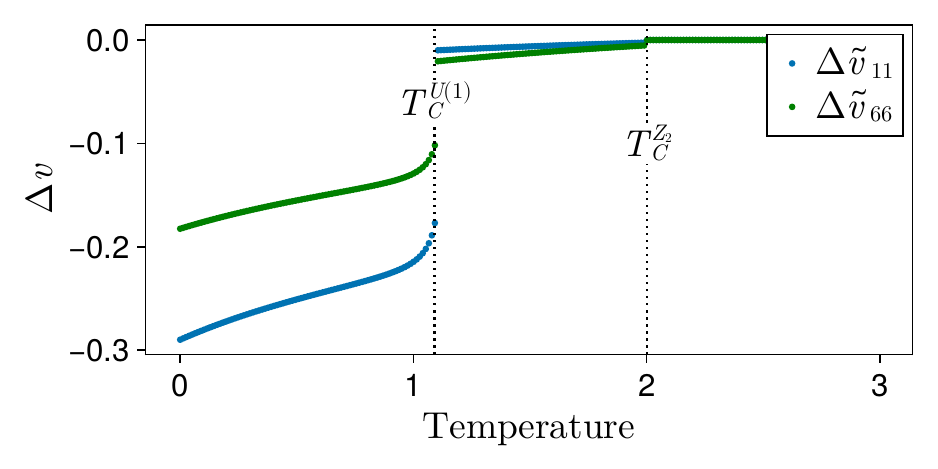}
    \caption{A typical ultrasound response for a model with nematicity in the $[110]$ direction.}
    \label{fig:nematic}
\end{figure}

We consider now the case that such nematicity of some origin is 
present in the quadrupling state. The symmetry breaking means we have to reanalyze the group and representation theory. Most significantly, the strain $u_{x,y}$ now transforms as $A_{1g}$ and can couple to the quadratic OP terms 
\begin{equation} \label{eq:OPquadterms}
|\psi_1|^2+|\psi_2|^2, |\psi_1|^2-|\psi_2|^2, \psi_1^\dagger\psi_2 + \psi_1\psi_2^\dagger, \Psi+\Psi^\dagger \, .
\end{equation}
The uniaxial strain $u_{x,x} - u_{y,y}$ still transforms as $B_{1g}$ and so 
cannot couple to the $s$-wave OPs.  The new free-energy term describing the coupling between strain and the OPs is
\begin{align}
\mathcal{F}_C^{s,s} &= \big(\gamma_1(|\psi_1|^2 + |\psi_2|^2) + \gamma_2( |\psi_1|^2 - |\psi_2|^2) +\gamma_5 |\Psi|^2  \nonumber \\
&+ \gamma_3(\psi_1\bar{\psi}_2 +  \bar{\psi}_1 \psi_2) +\gamma_4 (\Psi + \Psi^\dagger)   \big)(u_{x,x}+u_{y,y}) \, + \nonumber \\
&\big(\gamma_6(|\psi_1|^2 + |\psi_2|^2) + \gamma_5( |\psi_1|^2 - |\psi_2|^2 + \gamma_{10}|\Psi|^2) \nonumber \\
&+ \gamma_8(\psi_1\bar{\psi}_2 +  \bar{\psi}_1 \psi_2) +\gamma_9 (\Psi + \Psi^\dagger)  \big)u_{x,y} \, .  
\end{align}
We expect that $\gamma_{6,7,8,9,10}$ 
are small, proportional to the strength of the symmetry-breaking field. We will choose them to be one-half of the size of the corresponding coefficients $\gamma_{1,2,3,4,5}$. We can model the ultrasound response simply by modifying $\Gamma_{ij}$ in the formalism of Section \ref{sec:GLmodel}. We find that the new terms, with coefficients $\gamma_{6,7,8,9,10}$, create a non-zero response in both the longitudinal and $c_{66}$ sound modes at both critical temperatures, and no additional response in the transverse $B_{\rm 1g}$ mode, Fig.~\ref{fig:nematic}. 

Overall, a model with nematicity in the $[110]$ direction can explain the jump in $B_{2g}$ ultrasound mode for an $s$-wave OP. However, it also produces a small signal in the quadrupling phase. This signal is proportional to $\gamma_{9/10}$, which is controlled by the size of the nematicity. Hence, if the nematicity is small, the ultrasound response in the quadrupling phase may be small. However, the jump in $v_{66}$ at $T_{\rm c}^{U(1)}$ is also controlled by the size of the nematicity through $\gamma_{5,6,7}$.

We will now consider the case of an $s$-wave model with external stress. The detailed NMR studies have shown that the central line of the $^{75}As$ spectrum in this doping range has a double structure with broad satellite lines~\cite{Baertl2025}. The complementary analysis using scanning tunnelling microscopy has shown nanoscale clustering of the Ba/K atoms, which can explain the double structure of the central line and result in a local variation of the interatomic distances. However, the NMR spin lattice relaxation rate measurements on different parts of the spectra didn't show local variation in the superconducting transition temperature. Therefore, the impact of the clustering may be related to local symmetry-breaking strain. To understand the impact of the strain, we supposed that the system was externally stressed by, e.g., constant shear stresses $\sigma^0_{xy}$. This force transforms as $B_{2g}$ and its product with $u_{x,y}$ is invariant under all symmetry transformations. Hence, the product couples to any gauge-invariant functions of the order parameter. In detail, the terms
\begin{align}
&\sigma^0_{x,y} u_{x,y} \bigg( \alpha_1 (|\psi_1|^2+|\psi_2|^2) +\alpha_2(|\psi_1|^2-|\psi_2|^2)\nonumber  \\ &+\alpha_3\left(\psi_1\psi_2^\dagger + \psi_1^\dagger\psi_2\right) +  \alpha_4 (\Psi + \Psi^\dagger)  \bigg) \, 
\end{align}
where the $\alpha_i$ are coupling constants, should be added to the free energy. This can be modeled using the framework developed in Section \ref{sec:GLmodel} by updating the tensor $\Gamma_{ij}$.

In the decoupled limit, the newly added term gives an ultrasound response
\begin{equation}
\Delta \tilde{v}_T =\left(\sigma^0_{xy} \right)^2\left(\alpha_4^2D_4 - \frac{2\alpha_1^2}{b-2 c}-\frac{2 \alpha_2^2}{b+2 c}-\frac{\alpha_3^2}{8 c}  \right) \, ,
\end{equation}
corresponding to a linear response in the quadrupling phase and a jump at the superconducting transition. A similar calculation gives similar results for other external stresses. Hence, the presence of external stress can give rise to signals in all ultrasound components.

Now consider a model with higher-order strain coupling. All combinations of an $(s,s)$ order parameter transform as $A_{1g}$. These can couple to higher-order products of the strain tensor. The simplest are terms quadratic in strain. The possible terms, which affect strain in the plane, are
\begin{equation}
(u_{xx}+u_{yy})^2, (u_{xx}-u_{yy})^2,  u_{xy}^2.
\end{equation}
All three of these can couple to any quadratic term of the OP \eqref{eq:OPquadterms} considered earlier. We also consider terms of the form $|\Psi|^2$. So overall, there are fifteen terms of this kind. We can write the free-energy contribution of these terms in tensor notation as
\begin{equation}
F_{ijkl}(\psi,\Psi) u_{ij}u_{kl}.
\end{equation}

If we assume that the $c_{ijkl}$ defined in Eq.~\eqref{eq:tensornotation} are large, and hence $u_0$ is small, the formalism from Section \ref{sec:GLmodel} is only slightly modified. The dispersion relation Eq.~\eqref{eq:nonldisp} is modified by
\begin{equation}
    c_{ijkl} \to (c+F(\psi_0, \Psi_0))_{ijkl} \, .
\end{equation}
So, the sound velocity for the transverse sound wave in the normal state is simply
\begin{equation}
v^0_T = \sqrt{c_{11} - c_{12} + F_{1111} - F_{1122}} \, .
\end{equation}
The normalized change in $v_T$, in a large $c$ expansion, is given by
\begin{equation}
\frac{v_T-v^0_T}{v^0_T} \approx \frac{F_{1111} - F_{1122}}{2(c_{11} - c_{12})}.
\end{equation}
This expression can be written in terms of the order parameters, as follows
\begin{align} \label{eq:FFterms}
F_{1111} - &F_{1122} = f_1\left( |\psi_1|^2+|\psi_2|^2 \right) + f_2\left( |\psi_1|^2-|\psi_2|^2 \right) \nonumber \\
&f_3\left(\psi_1^\dagger\psi_2 + \psi_1\psi_2^\dagger\right) + f_4\left( \Psi+\Psi^\dagger\right) +f_5|\Psi|^2\, ,
\end{align}
with some new parameters $f_i$. So far, in this paper, we have modeled the phase transitions as being second order. Hence, the square of each order parameter grows approximately linearly with $T$ near $T_c$. As a result, the terms in Eq.~\eqref{eq:FFterms} are continuous across the phase transition: The new couplings generate a change in the slope of the ultrasound response. Hence, these terms cannot account for the discontinuous jump in the ultrasound data across the superconducting transition, when the phase transition is second order. However, it can account for the change in slope in the $c_{66}$ data from Fig.~\ref{fig:data2}b.

\begin{figure}
    \centering
    \includegraphics[width=\linewidth]{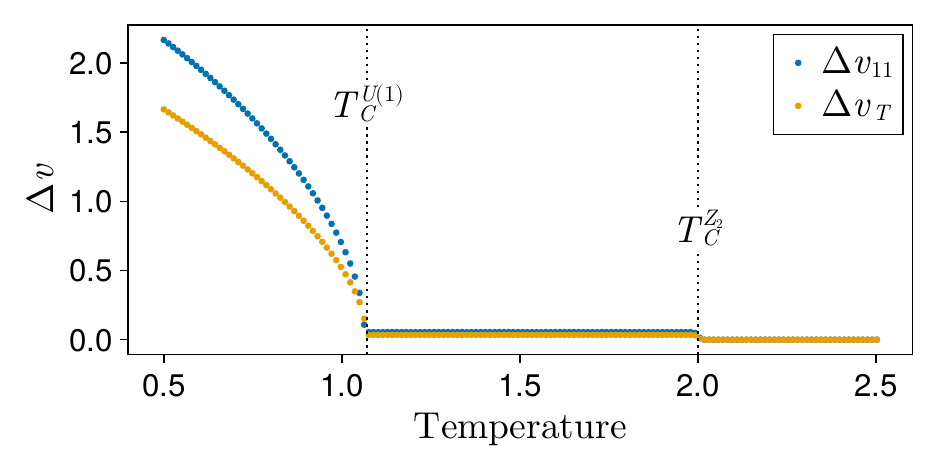}
    \caption{A typical ultrasound response at a first-order phase transition and higher-order strain coupling, for an $s$-wave model. There is a small non-zero response at the quadrupling phase transition and a jump at $T_c^{U(1)}$ in all sound modes.}
    \label{fig:first_order}
\end{figure}

We will now consider the case of a model with a first-order phase transition. In general, the phase transition from the quartic to the superconducting state can be first order when the quartic phase is not too large. This is seen in Monte-Carlo simulations of similar models for which, near the bicritical point, the phase transitions can be first order \cite{bojesen2014phase}. It was first pointed out and studied in detail in related models with different symmetry in Refs.~\cite{kuklov2004commensurate, kuklov2006deconfined}. Our simple model Eq.~\eqref{eq:free_energy} also contains a first-order phase transition from the quadrupling to BTRS superconducting phase when $A_i$, and hence $\Psi$, are small. We can model this using the parameters
\begin{align}
&(a(T), A_i, b, A_r, c, B_1, B_2,\gamma)   \\&= (1-T, \text{min}\left(0.02, 2-T\right), 1, 0.4, 0.2, 1, -0.1,-0.2) \, . \nonumber
\end{align}
One can check that the order parameters change discontinuously over $T=1$. We then also include the higher-order strain terms
\begin{align}
    &\left( |\psi_1|^2 + |\psi_2|^2\right)\left( f_1(u_{11} - u_{22} )^2 + g_1(u_{11} + u_{22} )^2 \right) \nonumber \\
    &+ |\Psi|^2\left( f_5(u_{11} - u_{22} )^2 + g_5( u_{11} + u_{22} )^2 \right) \, .
\end{align}
The ultrasound response is seen in Fig.~\ref{fig:first_order}. There are weak responses at the quadrupling phase transition (due to non-zero $g_5$ and $f_5$) and a discontinuity in the data at the superconducting transition. The discontinuities 
are due to the fact that the phase transition is first order. This relies on the fact that $A_i$ is small here. However, $A_i$ also controls the size of $\Psi_i^2$, and this controls the size of the response in the quadrupling phase. So, it seems difficult to construct a model of this kind with a large response in the quadrupling phase and a large jump at $T_c^{U(1)}$. Also note that the transitions in these models are very weakly first order \cite{bojesen2014phase,kuklov2006deconfined}.
These models do not contradict the experimental observations, since the existing calorimetry data cannot distinguish a second order transition from a  very weakly fist order transition~\cite{Grinenko2021state, shipulin2023calorimetric}.  

Finally, we will consider the case of derivative coupling. In the BTRS 
phase of \bkfa, there are spontaneous magnetic fields, whose values increase with decreasing temperature. These have been observed in the superconducting state at magic doping \cite{Grinenko2020superconductivity} and in the quadrupling state \cite{Grinenko2021state}.
For recent theoretical work on the origin of these fields, see \cite{Garaud2022effective}. The spontaneous magnetic fields imply persistent currents and, hence, the existence of stationary nonzero gradient terms. Nonzero gradient terms are important to describe muon spin rotation data of \bkfa, and hence, their potential role in the ultrasound response should be assessed. The allowed gradient terms depend on the OP symmetry, with a variety of consequences \cite{ benfenati2020magnetic, garaud2016thermoelectric}.

There are OP derivative terms which couple to the strain. One derivative term which couples to the $B_{1g}$ strain to an $s$-wave OP is
\begin{equation} \label{eq:dercouple}
  \left( | \mathcal{D}_x \Psi |^2 - | \mathcal{D}_y \Psi |^2 \right)\left(u_{xx} - u_{yy}\right) \, ,
\end{equation}
where $\mathcal{D}_i$ are the covariant derivatives. Such terms will only produce an ultrasound response where the order parameter is inhomogeneous, such as near defects, domain walls, and surfaces.
Microscale non-axially-symmetric defects lead to the appearance of spontaneous magnetic fields on relatively large scales in the simplest $s+is$ models \cite{garaud2014domain}.
Understanding this response, and whether it can be large enough to be seen in ultrasound experiments will require an 
elaboration on defect structures in the material and significant additional modeling.

\section{Discussion}

 Our main results are that the (i) ultrasound is sensitive to the phase transition in the electron quadrupling state.
(ii) The ultrasound experiments on \bkfa are
consistent with a time-reversal-symmetry breaking superconducting state. 

We have also 
{analyzed} how the ultrasound response depends on the symmetry of the electron quadrupling order parameters in more general setting, which will pave the way to ascertain the symmetry of the quadrupling phases in future works. The experimental data coincide best with our model of quadrupling order arising from a 
low-temperature $s+id_{xy}$ superconducting state, if we stay at the level of the simplest possible GL models and neglect that the expected small kink at $T_c^{Z_2}$ in the transverse mode was not reproducibly observed in the experiment. By contrast, 
the analysis of the polarization of spontaneous magnetic fields in the superconducting state \cite{Grinenko2020superconductivity} was more naturally explained by a model for which the low-temperature phase is an $s+is$-superconductor.
Nonetheless, we stress that our experimental data are inconsistent only with the simplest $s+is$ models. We show that multiple generalizations of $s+is$ models with additional inputs, such as explicit rotation-symmetry breaking by strain or defects in an $s+is$ state, can produce such an ultrasound response.
Hence, if one assumes $s+is$ superconductivity based on earlier experimental data \cite{Grinenko2020superconductivity}, then our ultrasound data may be indicative of the existence of strain and defects in the sample. Note that, similarly, detection of spontaneous magnetic field and spontaneous Nernst effects in basic $s+is$ models also require breaking spatial symmetry in addition to the breaking of time-reversal symmetry. Another promising approach is one with nematicity in the $[110]$ direction, which has some consistency with the enhanced nematic susceptibility seen in \cite{hong2022elastoresistivity}. The precise detail of the order parameter remains an intriguing question requiring a combination of further experimental and theoretical investigations.

Our overall conclusion is that the ultrasound data, although unable to uniquely determine the nature of the superconducting order parameter, does not contradict the fact that below the superconducting phase transition, \bkfa is a multicomponent superconductor that breaks $U(1)\times Z_2$ symmetry. This symmetry breaking is a necessary condition for the electron quadrupling transition at $T_c^{Z_2} > T_c^{U(1)}$. The multiple earlier experiments which studied $\mu$SR, spontaneous Nernst, and specific heat showed that the $Z_2$ symmetry is broken above the superconducting phase transition $T_c^{Z_2} > T_c^{U(1)}$~\cite{Grinenko2021state,shipulin2023calorimetric,Baertl2025}.  However, the expected weak anomaly at $T_c^{Z_2}$ was not yet observed reliably in the ultrasound experiments. For example, the previous experiment reported in Ref.~\cite{Grinenko2021state} indicates a kink-like feature at $T_c^{Z_2}$ in the transverse mode, but we have shown above (Fig.~\ref{fig:NPfit}) that this feature can be eliminated by modification of the phonon background. The new measurements at slightly lower doping levels did not reveal, within the experimental errors, any feature in the $c_{66}$ mode at $T_c^{Z_2}$. On the other hand, there is a possible kink in the longitudinal mode (Fig.~\ref{fig:data1}b) that is qualitatively consistent with the theoretical expectations for the $s+is$ scenario (Fig.~\ref{fig:US_response_fav}), but this feature was not resolved in the previous experiments~\cite{Grinenko2021state}. The difficulties in resolving anomalies or kinks at $T_c^{Z_2}$ are attributed to the large experimental errors related to the small thickness of the crystals and local strain possibly caused by Ba/K nano-scale clustering. Hence, given experimental errors and the small size of the expected anomaly at $T_c^{Z_2}$, the overall ultrasound data support the existence of an electron quadrupling phase with ${Z_2}$ broken symmetry above the superconducting phase transition.  
We anticipate that the future availability of large, high-quality crystals will enable a conclusive investigation of possible weaker anomalies at the $T_c^{Z_2}$ transition.

\section{Methods}

\subsection{Samples}

Plate-like \bkfa single-crystals were examined by X-ray diffraction. The $c$-axis lattice parameters were calculated from the X-ray diffraction data using the Nelson–Riley function. The doping level $x$ of K for the single crystals was determined using the relation between the c-axis lattice parameter and the K doping obtained in previous studies~\cite{Kihou2016}. 

\subsection{The orientation of the single-crystals}

The orientation was verified using polarized Raman spectroscopy. The tetagoanl \bkfa system has one $B_{1g}$-symmetry Raman-active phonon, and it does not have any $B_{2g}$-symmetry phonons. In polarized Raman data, taken with incident/scattered light cross-polarization, the $B_{1g}$ phonon appears for incident = [110], scattered = [-110] (X’Y’) light polarization geometry while $B_{2g}$-symmetry excitations corresponding to the incident = [100], scattered = [-010] (XY) geometry do not contain the phonon.

\subsection{Ultrasound measurements}

The measurements were performed using a pulse-echo phase-sensitive detection technique~\cite{Zherlitsyn_2014} in a gas-flow cryostat. A pair of piezoelectric LiNbO$_3$ resonance transducers were glued to parallel opposite (100) crystal surfaces to generate and detect acoustic waves. We used Z- and X-cut transducers (Boston Piezo-Optics) with fundamental frequencies close to 30 MHz. For further details, see Ref.~\cite{Grinenko2021state}.

\section{Data Availability}
The experimental data shown in Figs. 2 and 3 are available on request.

\section{Acknowledgements}
We thank Connor Garrity and Girsh Blumberg for orienting the single crystals using polarized Raman spectroscopy and for discussions.
CH is supported by the Carl Trygger Foundation through the grant CTS 20:25.
EB was supported by the Swedish Research Council Grants  2022-04763, by Olle Engkvists Stiftelse, and partially by the Wallenberg Initiative Materials Science
for Sustainability (WISE) funded by the Knut and Alice Wallenberg
Foundation. YW and VG are supported by the NSFC grants 12374139 and 12350610235. 
We acknowledge the support of the HLD at HZDR, a member of the European
Magnetic Field Laboratory (EMFL), and the Würzburg-Dresden Cluster of Excellence on Complexity and Topology in Quantum Matter--ct.qmat (EXC 2147, Project No. 390858490).

\section{Author contribution}

C.H. Developed  the theory, performed calculations, theoretical data analysis and wrote the paper;
I.S. performed ultrasound measurements, analysed the data; F.C. performed electrical and thermoelectrical transport experiments; Y.L. performed magnetic susceptibility measurements; J.W. supervised research at HLD-EMFL; H.H.K. performed ultrasound measurements, supervised research at TUD; S.Z. performed ultrasound experiments;  V.G. initiated the project, performed ultrasound experiments, analysed the data and wrote the paper.
E.B.  Contributed  model building,  contributed to theoretical analysis.  interpretation of the results, and contributed to writing the paper.

\section{Competing Interests}
The authors declare no competing interests.
 
\bibliographystyle{apsrev4-1}
\bibliography{mainbib.bib}

@Article{Grinenko2021state,
  author      = {Grinenko, V. and Weston, D. and Caglieris, F. and Wuttke, C. and Hess, C. and Gottschall, T. and Maccari, I. and Gorbunov, D. and Zherlitsyn, S. and Wosnitza, J. and Rydh, A. and Kihou, K. and Lee, C.-H. and Sarkar, R. and Dengre, S. and Garaud, J. and Charnukha, A. and H\"uhne, R. and Nielsch, K. and B\"uchner, B. and Klauss, H.-H. and Babaev, E.},
  title       = {State with spontaneously broken time-reversal symmetry above the superconducting phase transition},
  journal     = {Nat. Phys.},
  year        = {2021},
  number   = {17},
  pages    = {1254–1259},
  volume   = {},
  comment  = {},
  doi      = {10.1038/s41567-021-01350-9},
}

@article{maccari2022effects,
  title={Effects of intercomponent couplings on the appearance of time-reversal symmetry breaking fermion-quadrupling states in two-component london models},
  author={Maccari, Ilaria and Babaev, Egor},
  journal={Physical Review B},
  volume={105},
  number={21},
  pages={214520},
  year={2022},
  publisher={APS}
}

@Article{Bojesen2013time,
  author    = {Bojesen, Troels Arnfred and Babaev, Egor and Sudb\o{}, Asle},
  journal   = {Phys. Rev. B},
  title     = {Time reversal symmetry breakdown in normal and superconducting states in frustrated three-band systems},
  year      = {2013},
  month     = {Dec},
  pages     = {220511},
  volume    = {88},
  comment   = {Multicomponent Ginzburg-Landau Monte Carlo. TRSB project},
  doi       = {10.1103/PhysRevB.88.220511},
  issue     = {22},
  numpages  = {4},
  publisher = {American Physical Society},
}

@article{bojesen2014phase,
  author    = {Bojesen, Troels Arnfred and Babaev, Egor and Sudb\o{}, Asle},
  journal   = {Phys. Rev. B},
  title     = {Phase transitions and anomalous normal state in superconductors with broken time-reversal symmetry},
  year      = {2014},
  month     = {Mar},
  pages     = {104509},
  volume    = {89},
  comment   = {Multicomponent London Wang-Landau. TRSB project},
  doi       = {10.1103/PhysRevB.89.104509},
  issue     = {10},
  numpages  = {11},
  publisher = {American Physical Society},
}

@article{shipulin2023calorimetric,
  title={Calorimetric evidence for two phase transitions in Ba$_{1-x}$K$_x$Fe$_2$As$_2$ with fermion pairing and quadrupling states},
  author={Shipulin, Ilya and Stegani, Nadia and Maccari, Ilaria and Kihou, Kunihiro and Lee, Chul-Ho and Li, Yongwei and H{\"u}hne, Ruben and Klauss, Hans-Henning and Putti, Marina and Caglieris, Federico and others},
journal = {Nature Communications},
  volume = {14},
  pages = {6734},
  year={2023},
doi={10.1038/s41467-023-42459-0}
}

@misc{zheng2024observation,
      title={Observation of counterflow superfluidity in a two-component Mott insulator}, 
      author={Yong-Guang Zheng and An Luo and Ying-Chao Shen and Ming-Gen He and Zi-Hang Zhu and Ying Liu and Wei-Yong Zhang and Hui Sun and Youjin Deng and Zhen-Sheng Yuan and Jian-Wei Pan},
      year={2024},
      eprint={2403.03479},
      archivePrefix={arXiv},
      primaryClass={cond-mat.quant-gas}
}

@article{babaev2004superconductor,
	Author = {Babaev, Egor and Sudb{\o}, Asle and Ashcroft, NW},
	Journal = {Nature},
	Number = {7009},
	Pages = {666},
	Publisher = {Nature Publishing Group},
	Title = {A superconductor to superfluid phase transition in liquid metallic hydrogen},
	Volume = {431},
	Year = {2004},
doi={10.1038/nature02910}}

@Article{Babaev2004phase,
  author   = {Egor Babaev},
  journal  = {Nucl. Phys. B},
  title    = {Phase diagram of planar $U(1)\times U(1)$ superconductor: Condensation of vortices with fractional flux and a superfluid state},
  year     = {2004},
  issn     = {0550-3213},
  number   = {3},
  pages    = {397 - 412},
  volume   = {686},
  doi      = {https://doi.org/10.1016/j.nuclphysb.2004.02.021},
}

@Article{Smiseth2005field,
  author   = {{Smiseth}, J. and {Sm{\o}rgrav}, E. and {Babaev}, E. and {Sudb{\o}}, A.},
  journal  = {Phys. Rev. B},
  title    = {Field- and temperature-induced topological phase transitions in the three-dimensional $N$-component London superconductor},
  year     = {2005},
  month    = jun,
  number   = {21},
  pages    = {214509},
  volume   = {71},
  adsurl   = {http://adsabs.harvard.edu/abs/2005PhRvB..71u4509S},
  doi      = {10.1103/PhysRevB.71.214509},
  eid      = {214509},
}

@Article{Kuklov2008deconfined,
  author    = {Kuklov, A. B. and Matsumoto, M. and Prokof'ev, N. V. and Svistunov, B. V. and Troyer, M.},
  journal   = {Phys. Rev. Lett.},
  title     = {Deconfined Criticality: Generic First-Order Transition in the SU(2) Symmetry Case},
  year      = {2008},
  month     = {Aug},
  pages     = {050405},
  volume    = {101},
  comment   = {flowgram},
  doi       = {10.1103/PhysRevLett.101.050405},
  issue     = {5},
  numpages  = {4},
  publisher = {American Physical Society},
}

@Article{Herland2010phase,
  author    = {Herland, Egil V. and Babaev, Egor and Sudb\o{}, Asle},
  journal   = {Phys. Rev. B},
  title     = {Phase transitions in a three dimensional ${U}(1)\ifmmode\times\else\texttimes\fi{}{U}(1)$ lattice London superconductor: Metallic superfluid and charge-$4e$ superconducting states},
  year      = {2010},
  month     = {Oct},
  pages     = {134511},
  volume    = {82},
  comment   = {Good starting point for reading about multicomponent Ginzburg-Landau Monte Carlo. Helicity modulus.},
  doi       = {10.1103/PhysRevB.82.134511},
  issue     = {13},
  numpages  = {16},
  publisher = {American Physical Society},
}

@Book{Svistunov2015,
  author    = {Boris Svistunov and Egor Babaev and Nikolay Prokofev},
  publisher = {CRC Press},
  title     = {Superfluid States of Matter},
  year      = {2015},
  comment   = {Hi Wenlong and Daniel The descriptoon of Faddevv-Skyrme model can be found in chapter 6 of our book All the best Egor},
}

@article{agterberg2008dislocations,
  title={Dislocations and vortices in pair-density-wave superconductors},
  author={Agterberg, DF and Tsunetsugu, H},
  journal={Nature Physics},
  volume={4},
  number={8},
  pages={639--642},
  year={2008},
  publisher={Nature Publishing Group},
doi={10.1038/nphys999}
}

@article{berg2009charge,
  title={Charge-4e superconductivity from pair-density-wave order in certain high-temperature superconductors},
  author={Berg, Erez and Fradkin, Eduardo and Kivelson, Steven A},
  journal={Nature Physics},
  volume={5},
  number={11},
  pages={830--833},
  year={2009},
  publisher={Nature Publishing Group},
doi={10.1038/nphys1389}
}

@article{Radzihovsky2009liquid,
  title = {Quantum Liquid Crystals in an Imbalanced Fermi Gas: Fluctuations and Fractional Vortices in Larkin-Ovchinnikov States},
  author = {Radzihovsky, Leo and Vishwanath, Ashvin},
  journal = {Phys. Rev. Lett.},
  volume = {103},
  issue = {1},
  pages = {010404},
  numpages = {4},
  year = {2009},
  month = {Jul},
  publisher = {American Physical Society},
  doi = {10.1103/PhysRevLett.103.010404},
  url = {https://link.aps.org/doi/10.1103/PhysRevLett.103.010404}
}

@Article{Iguchi2023,
  author  = {Iguchi, Y. and Shi, R. and Kihou, K and Lee, C.H. and Barkman, M. and Benfenat, A.L. and  Grinenko, V. and Babaev, E. and Moler, K.A.},
  title   = {Superconducting vortices carrying a temperature-dependent fraction of the flux quantum},
  journal = {Science},
  year    = {2023},
  volume  = {380},
  number  = {},
  pages   = {1244-1247},
  doi     = {10.1126/science.abp9979},
  eprint  = {},
}

@article{hebel1957nuclear,
  title={Nuclear relaxation in superconducting aluminum},
  author={Hebel, LC and Slichter, CP},
  journal={Physical Review},
  volume={107},
  number={3},
  pages={901},
  year={1957},
  publisher={APS},
doi={https://doi.org/10.1103/PhysRev.107.901}
}

@article{golding1985observation,
  title={Observation of a Collective Mode in Superconducting U Be 13},
  author={Golding, Brage and Bishop, DJ and Batlogg, B and Haemmerle, WH and Fisk, Z and Smith, JL and Ott, HR},
  journal={Physical review letters},
  volume={55},
  number={22},
  pages={2479},
  year={1985},
  publisher={APS},
doi={https://doi.org/10.1103/PhysRevLett.55.2479}
}

@article{muller1986observation,
  title={Observation of a lambda-shaped ultrasonic attenuation peak in superconducting UPt3},
  author={M{\"u}ller, V and Maurer, D and Scheidt, Ernst-Wilhelm and Roth, Ch and L{\"u}ders, K and Bucher, E and B{\"o}mmel, HE},
  journal={Solid state communications},
  volume={57},
  number={5},
  pages={319--321},
  year={1986},
  publisher={Elsevier},
doi={https://doi.org/10.1016/0038-1098(86)90099-2}
}

@book{tinkham2004introduction,
  title={Introduction to superconductivity},
  author={Tinkham, Michael},
  year={2004},
  publisher={Courier Corporation}
}

@article{ghosh2020one,
  title={One-component order parameter in URu2Si2 uncovered by resonant ultrasound spectroscopy and machine learning},
  author={Ghosh, Sayak and Matty, Michael and Baumbach, Ryan and Bauer, Eric D and Shekhter, Arkady and Mydosh, JA and Kim, Eun-Ah and Ramshaw, BJ},
  journal={Science advances},
  volume={6},
  number={10},
  pages={eaaz4074},
  year={2020},
  publisher={American Association for the Advancement of Science},
doi={10.1126/sciadv.aaz4074}
}

@Article{Benhabib2021,
	author   = {S. Benhabib and C. Lupien and I. Paul and L. Berges and M. Dion and M. Nardone and A. Zitouni and Z. Q. Mao and Y. Maeno and A. Georges and L. Taillefer and C. Proust},
	title    = {Ultrasound evidence for a two-component
superconducting order parameter in Sr$_2$RuO$_4$},
	journal  = {Nature Physics},
	year     = {2021},
	volume   = {17},
	number   = {},
	pages    = {194–198},
	doi      = {10.1038/s41567-020-1033-3},
	eprint   = {},
}

@article{sigrist2002ehrenfest,
  title={Ehrenfest relations for ultrasound absorption in Sr$_2$RuO$_4$},
  author={Sigrist, Manfred},
  journal={Progress of Theoretical Physics},
  volume={107},
  number={5},
  pages={917--925},
  year={2002},
  publisher={Oxford University Press},
doi={10.1143/PTP.107.917}
}

@article{ghosh2021thermodynamic,
  title={Thermodynamic evidence for a two-component superconducting order parameter in Sr2RuO4},
  author={Ghosh, Sayak and Shekhter, Arkady and Jerzembeck, F and Kikugawa, N and Sokolov, Dmitry A and Brando, Manuel and Mackenzie, AP and Hicks, Clifford W and Ramshaw, BJ},
  journal={Nature Physics},
  volume={17},
  number={2},
  pages={199--204},
  year={2021},
  publisher={Nature Publishing Group UK London},
doi={https://doi.org/10.1038/s41567-020-1032-4}
}

@article{hong2022elastoresistivity,
  title={Elastoresistivity of Heavily Hole-Doped 122 Iron Pnictide Superconductors},
  author={Hong, Xiaochen and Sykora, Steffen and Caglieris, Federico and Behnami, Mahdi and Morozov, Igor and Aswartham, Saicharan and Grinenko, Vadim and Kihou, Kunihiro and Lee, Chul-Ho and B{\"u}chner, Bernd and others},
  journal={Frontiers in Physics},
  volume={10},
  pages={853717},
  year={2022},
  publisher={Frontiers},
doi={10.3389/fphy.2022.853717}
}

@Article{Kuklov2006decoinfined,
  author   = {A.B. Kuklov and N.V. Prokof'ev and B.V. Svistunov and M. Troyer},
  journal  = {Ann. Phys.},
  title    = {Deconfined criticality, runaway flow in the two-component scalar electrodynamics and weak first-order superfluid-solid transitions},
  year     = {2006},
  issn     = {0003-4916},
  note     = {July 2006 Special Issue},
  number   = {7},
  pages    = {1602 - 1621},
  volume   = {321},
  comment  = {flowgram},
  doi      = {https://doi.org/10.1016/j.aop.2006.04.007},
}

@book{luthi2007physical,
  title={Physical acoustics in the solid state},
  author={L{\"u}thi, Bruno},
  volume={148},
  year={2007},
  publisher={Springer Science \& Business Media}
}

@article{benfenati2020magnetic,
  title={Magnetic signatures of domain walls in s+ i s and s+ i d superconductors: Observability and what that can tell us about the superconducting order parameter},
  author={Benfenati, Andrea and Barkman, Mats and Winyard, Thomas and Wormald, Alex and Speight, Martin and Babaev, Egor},
  journal={Physical Review B},
  volume={101},
  number={5},
  pages={054507},
  year={2020},
  publisher={APS},
doi={10.1103/PhysRevB.101.054507}
}

@article{reid2012universal,
  title={Universal heat conduction in the iron arsenide superconductor KFe$_2$As$_2$: Evidence of a d-wave state},
  author={Reid, J-Ph and Tanatar, Makariy A and Juneau-Fecteau, A and Gordon, RT and de Cotret, S Ren{\'e} and Doiron-Leyraud, N and Saito, T and Fukazawa, H and Kohori, Y and Kihou, K and others},
  journal={Physical Review Letters},
  volume={109},
  number={8},
  pages={087001},
  year={2012},
  publisher={APS},
doi={10.1103/PhysRevLett.109.087001}
}

@article{abdel2013evidence,
  title={Evidence of d-wave superconductivity in K 1- x Na x Fe 2 As 2 (x= 0, 0.1) single crystals from low-temperature specific-heat measurements},
  author={Abdel-Hafiez, M and Grinenko, V and Aswartham, S and Morozov, I and Roslova, M and Vakaliuk, O and Johnston, S and Efremov, DV and Van Den Brink, J and Rosner, H and others},
  journal={Physical Review B},
  volume={87},
  number={18},
  pages={180507},
  year={2013},
  publisher={APS},
doi={10.1103/PhysRevB.87.180507}
}

@article{wu2022nodal,
  title={Nodal $s_\pm$ Pairing Symmetry in an Iron-Based Superconductor with only Hole Pockets},
  author={Wu, Dingsong and Jia, Junjie and Yang, Jiangang and Hong, Wenshan and Shu, Yingjie and Miao, Taimin and Yan, Hongtao and Rong, Hongtao and Ai, Ping and Zhang, Xing and others},
  journal={Nat. Phys.},
volume={20},
  number={},
  pages={571–578},
   publisher={Nature},
  year={2024},
doi={10.1038/s41567-023-02348-1}
}

@article{richard2009angle,
  title={Angle-resolved photoemission spectroscopy of the Fe-based Ba$_{0.6}$K$_{0.4}$Fe$_2$As$_2$ high temperature superconductor: evidence for an orbital selective electron-mode coupling},
  author={Richard, P and Sato, T and Nakayama, K and Souma, S and Takahashi, T and Xu, Y-M and Chen, GF and Luo, JL and Wang, NL and Ding, H},
  journal={Physical review letters},
  volume={102},
  number={4},
  pages={047003},
  year={2009},
  publisher={APS},
doi={10.1103/PhysRevLett.102.047003}
}

@Article{Cai2021,
  author  = {Cai, Y. and Huang, J. and Miao, T. and Wu, D. and Gao, Q. and Li, C. and Xu, Y. and Jia, J. and Wang, Q. and Huang, Y. and Liu, G. and Zhang, F. and Zhang, S. and Yang, F. and Wang, Z. and Peng, Q. and Xu, Z. and Zhao, L. and Zhou X.},
  title   = {Genuine electronic structure and superconducting gap structure in (Ba$_{0.6}$K$_{0.4}$)Fe$_2$As$_2$ superconductor},
  journal = {Science Bulletin},
  year    = {2021},
  volume  = {66},
  number  = {},
  pages   = {1839-1848},
  doi     = {10.1016/j.scib.2021.05.015},
  eprint  = {},
}

@article{luo2009quasiparticle,
  title={Quasiparticle heat transport in single-crystalline Ba$_{1-x}$K$_x$Fe$_2$As$_2$: Evidence for a k-dependent superconducting gap without nodes},
  author={Luo, XG and Tanatar, MA and Reid, J-Ph and Shakeripour, H and Doiron-Leyraud, N and Ni, Ni and Bud’ko, Sergey L and Canfield, PC and Luo, Huiqian and Wang, Zhaosheng and others},
  journal={Physical Review B},
  volume={80},
  number={14},
  pages={140503},
  year={2009},
  publisher={APS},
doi={10.1103/PhysRevB.80.140503}
}

@article{reid2016doping,
  title={Doping evolution of the superconducting gap structure in the underdoped iron arsenide Ba$_{1-x}$K$_x$Fe$_2$As$_2$ revealed by thermal conductivity},
  author={Reid, J-Ph and Tanatar, MA and Luo, XG and Shakeripour, H and de Cotret, S Ren{\'e} and Juneau-Fecteau, A and Chang, J and Shen, B and Wen, H-H and Kim, H and others},
  journal={Physical Review B},
  volume={93},
  number={21},
  pages={214519},
  year={2016},
  publisher={APS},
doi={10.1103/PhysRevB.93.214519}
}

@article{kuklov2004commensurate,
  title={Commensurate two-component bosons in an optical lattice: Ground state phase diagram},
  author={Kuklov, Anatoly and Prokof’ev, Nikolay and Svistunov, Boris},
  journal={Physical review letters},
  volume={92},
  number={5},
  pages={050402},
  year={2004},
  publisher={APS},
doi={10.1103/PhysRevLett.92.050402}
}

@Article{Grinenko2020superconductivity,
  author   = {V. Grinenko and R. Sarkar and K. Kihou and C. H. Lee and I. Morozov and S. Aswartham and B. B{\"u}chner and P. Chekhonin and W. Skrotzki and K. Nenkov and R. H{\"u}hne and K. Nielsch and S. -L. Drechsler and V. L. Vadimov and M. A. Silaev and P. Volkov and I. Eremin and H. Luetkens and H. H. Klauss},
  journal  = {Nat. Phys.},
  title    = {Superconductivity with broken time-reversal symmetry inside a superconducting $s$-wave state},
  year     = {2020},
  pages    = {789–794},
  volume   = {16},
  comment  = {TRSB project},
doi={10.1038/s41567-020-0886-9}
}

@Article{Garaud2022effective,
  author      = {Garaud, J. and Babaev, E.},
  title       = {Effective Model and Magnetic Properties of the Resistive Electron Quadrupling State},
  journal     = {Phys. Rev. Lett.},
  year        = {2022},
  number   = {129},
  pages    = {087602},
  volume   = {},
  comment  = {},
  doi      = {10.1103/PhysRevLett.129.087602},
}

@article{garaud2014domain,
  title={Domain Walls and Their Experimental Signatures in s+ i s Superconductors},
  author={Garaud, Julien and Babaev, Egor},
  journal={Phys. Rev. Lett.},
  volume={112},
  number={1},
  pages={017003},
  year={2014},
  publisher={APS},
doi={https://doi.org/10.1103/PhysRevLett.112.017003}
}

@article{garaud2016thermoelectric,
  title={Thermoelectric signatures of time-reversal symmetry breaking states in multiband superconductors},
  author={Garaud, Julien and Silaev, Mihail and Babaev, Egor},
  journal={Phys. Rev. Lett.},
  volume={116},
  number={9},
  pages={097002},
  year={2016},
  publisher={APS},
doi={https://doi.org/10.1103/PhysRevLett.116.097002}
}

@article{kuklov2006deconfined,
  title={Deconfined criticality, runaway flow in the two-component scalar electrodynamics and weak first-order superfluid-solid transitions},
  author={Kuklov, AB and Prokof’Ev, NV and Svistunov, BV and Troyer, Matthias},
  journal={Annals of Physics},
  volume={321},
  number={7},
  pages={1602--1621},
  year={2006},
  publisher={Elsevier},
doi = {https://doi.org/10.1016/j.aop.2006.04.007}
}

@Article{Grinenko2021split,
	author   = {Grinenko, Vadim and Ghosh, Shreenanda and Sarkar, Rajib and Orain, Jean-Christophe and Nikitin, Artem  and Elender, Matthias and Das, Debarchan  and Guguchia, Zurab and Bruckner, Felix and  Barber, Mark E. and Park, Joonbum and Kikugawa, Naoki and Sokolov, Dmitry A. and Bobowski, Jake S. and Miyoshi, Takuto and Maeno, Yoshiteru and Mackenzie, Andrew P. and Luetkens, Hubertus and Hicks, Clifford W. and Klauss, Hans-Henning},
	title    = {Split superconducting and time-reversal symmetry-breaking transitions in  {S}r$_2${R}u{O}$_4$ under stress},
	journal  = {Nat. Phys.},
	year     = {2021},
	volume   = {},
	number   = {},
	pages    = {},
	doi      = {10.1038/s41567-021-01182-7},
}

@Article{Grinenko2017bkfa,
  author    = {Grinenko, V. and Materne, P. and Sarkar, R. and Luetkens, H. and Kihou, K. and Lee, C. H. and Akhmadaliev, S. and Efremov, D. V. and Drechsler, S.-L. and Klauss, H.-H.},
  title     = {Superconductivity with broken time-reversal symmetry in ion-irradiated {B}a$_{027}${K}$_{0.73}${F}e$_2${A}$_2$ single crystals},
  journal   = {Phys. Rev. B},
  year      = {2017},
  volume    = {95},
  pages     = {214511},
  month     = {Jun},
  doi       = {10.1103/PhysRevB.95.214511},
  issue     = {21},
  numpages  = {5},
  publisher = {American Physical Society},
}

@Article{Baertl2025,
  author  = {B\"artl, F. and Stegani, N. and Caglieris, F. and Shipulin, I. and Li Y. and Hu, Q. and Zheng, Z. and Yim, C.M. and Luther, S. and Wosnitza, W. and Sarkar, R. and Klauss, H.-H. and Garaud, J. and  Babaev, E. and K\"uhne, H. and Grinenko, V.},
  title   = {Evidence of pseudogap and absence of spin magnetism in the time-reversal-symmetry-breaking state of {B}a$_{1−x}${K}$_x${F}e$_2${A}s$_2$},
  journal = {arXiv:2501.11936},
  year    = {2025},
  volume  = {},
  number  = {},
  pages   = {},
  doi     = {doi.org/10.48550/arXiv.2501.11936},
  eprint  = {},
}

@Article{Kihou2016,
  author  = {Kihou ,Kunihiro and Saito ,Taku and Fujita ,Kay and Ishida ,Shigeyuki and Nakajima ,Masamichi and Horigane ,Kazumasa and Fukazawa ,Hideto and Kohori ,Yoh and Uchida ,Shin-ichi and Akimitsu ,Jun and Iyo ,Akira and Lee ,Chul-Ho and Eisaki ,Hiroshi},
  title   = {Single-Crystal Growth of Ba1-xKxFe2As2 by KAs Self-Flux Method},
  journal = {J. Phys. Soc. Jpn.},
  year    = {2016},
  volume  = {85},
  number  = {3},
  pages   = {034718},
  doi     = {10.7566/JPSJ.85.034718},
  eprint  = {https://doi.org/10.7566/JPSJ.85.034718},
}

@Article{Zherlitsyn_2014,
	author   = {S. Zherlitsyn and S. Yasin and J. Wosnitza and A. A. Zvyagin and A. V. Andreev and V. Tsurkan},
	title    = {Spin-lattice effects in selected antiferromagnetic materials},
	journal  = {Low Temp. Phys.},
	year     = {2014},
	volume   = {40},
	number   = {},
	pages    = {123},
	doi      = {},
	eprint   = {},
}

\clearpage

\begin{table}
    \centering
     \caption{The product table for the irreps of $D_{4h}$. We have disregarded antisymmetric elements.}
        \begin{tabular}{c|ccccc}
                 $\otimes$&  $A_{1g}$& $A_{2g}$ & $B_{1g}$ & $B_{2g}$ & $E_g $\\ \hline
       $A_{1g}$  & $A_{1g}$ & $A_{2g}$ & $B_{1g}$ & $\boldsymbol{B_{2g}}$ & $E_g $ \\
       $A_{2g}$  &  & $A_{1g}$ & $\boldsymbol{B_{2g}}$ & $B_{1g}$ & $E_g$ \\
        $B_{1g}$  &  &  & $A_{1g}$ & $A_{2g}$ & $E_g$ \\
       $B_{2g}$  &  &  &  & $A_{1g}$ & $E_g$ \\
        $E_g $ &  &  &  &  & $A_{1g} \oplus B_{1g} \oplus \boldsymbol{B_{2g}}$  \\
    \end{tabular}
    \label{tab:producttable}
\end{table}

\end{document}